\documentclass[pra,twocolumn,superscriptaddress]{revtex4}
\usepackage{graphicx}
\usepackage{amsmath}

\newcounter{defcounter}
\begin{document}
 
\title{Local protein solvation drives direct down-conversion in phycobiliprotein PC645 via incoherent vibronic transport}

\author{Samuel M. Blau}
\thanks{These authors contributed equally to this work.}
\affiliation{Department of Chemistry and Chemical Biology, Harvard University, 12 Oxford St, Cambridge, Massachusetts 02138, USA}
\author{Doran I. G. Bennett}
\thanks{These authors contributed equally to this work.}
\affiliation{Department of Chemistry and Chemical Biology, Harvard University, 12 Oxford St, Cambridge, Massachusetts 02138, USA}
\author{Christoph Kreisbeck}
\affiliation{Department of Chemistry and Chemical Biology, Harvard University, 12 Oxford St, Cambridge, Massachusetts 02138, USA}
\author{Gregory D. Scholes}
\affiliation{Department of Chemistry, Princeton University, Princeton, NJ 08544, USA}
\affiliation{Senior Fellow. Canadian Institute for Advanced Research, Bioinspired Solar Energy Program, Toronto, ON M5G 1Z8, Canada}
\author{Al\'{a}n Aspuru-Guzik}
\email{aspuru@chemistry.harvard.edu}
\affiliation{Department of Chemistry and Chemical Biology, Harvard University, 12 Oxford St, Cambridge, Massachusetts 02138, USA}
\affiliation{Senior Fellow. Canadian Institute for Advanced Research, Bioinspired Solar Energy Program, Toronto, ON M5G 1Z8, Canada}

\begin{abstract}
Mechanisms controlling excitation energy transport (EET) in light-harvesting complexes remain controversial. Following the observation of long-lived beats in two-dimensional electronic spectroscopy of PC645, vibronic coherence, the delocalization of excited states between pigments supported by a resonant vibration, has been proposed to enable direct down-conversion from the highest-energy states to the lowest-energy pigments. Here, we instead show that for phycobiliprotein PC645 an incoherent vibronic transport mechanism is at play. We quantify the solvation dynamics of individual pigments using ab initio QM/MM nuclear dynamics. Our atomistic spectral densities reproduce experimental observations ranging from absorption and fluorescence spectra to the timescales and selectivity of down-conversion observed in transient absorption measurements. We demonstrate that bilin solvation controls EET pathways and that direct down-conversion proceeds incoherently, enhanced by large reorganization energies and a broad collection of high-frequency vibrations. We thus suggest that engineering local solvation dynamics represents a potential design principle for nanoscale control of EET.
\end{abstract}
\maketitle

\section*{Introduction}
\hspace{\parindent} Understanding the solvation dynamics of multiple pigments in a heterogeneous medium is essential for being able to control excited state dynamics in artificial materials. Natural photosynthesis demonstrates this capability in light-harvesting complexes, which exert nanoscale control over excitation transport between pigments via the protein environment. However, at present, it is not possible to experimentally disentangle excitation dynamics from the many timescales of the vibrational bath, which include intramolecular pigment vibrations, fluctuations of neighboring amino acids, reorientation of proximate ‘biological’ water, and reorganization of the bulk water solvent. Thus, quantitatively connecting the atomic scale motions of the pigment-protein environment to the dynamics of excitation transport in light-harvesting complexes remains a grand challenge at the interface of materials science and chemical physics.

It has long been speculated that simple design principles connect the complex atomic structure of pigment environments to the regulation of EET pathways in LHCs \cite{Alan2016, Scholes:2011iq, Caram:2012cc,Kruger:2016cs,Muh:2007tb}. Recently, there has been growing interest in understanding the role of the vibrational environment in enhancing transport between energetically detuned pigments. The presence of a vibration with the same frequency as the pigment energy gap can enhance transport either coherently, via delocalization, or incoherently, via discrete hopping. Long-lived beat signals in nonlinear spectroscopic measurements of multiple LHCs have been interpreted as evidence for vibronic coherence, the delocalization of excited states between pigments (excitons) supported by a long-lived resonant vibration \cite{Novoderezhkin:2017jt, Fuller:2014iz, Romero:2014jm, Dean2016a, Christensson:2012gp, Chin:2013ia,  Kreisbeck2012a, Kolli2012a, Perlik:2015bc, Novelli:2015es}. To date, vibronic coherence has been proposed to enhance transport between energetically remote states in two excitation transport processes associated with light-harvesting: charge separation in the reaction center \cite{Novoderezhkin:2017jt, Fuller:2014iz, Romero:2014jm} and direct down-conversion in the bilin-containing protein PC645 \cite{Dean2016a}. Here, we instead show that for phycobiliprotein PC645 an incoherent vibronic transport mechanism is at play. We find that flexible bilins, unlike rigid chlorophylls, exhibit large reorganization energies tuned by individual bilin-protein environments, suggesting that phycobiliproteins undergo incoherent energy transfer enhanced by the presence of a broad collection of high-frequency, intramolecular vibrations.

{\it{In vivo}}, PC645 absorbs high-energy photons and down-converts the excitation energy to be resonant with the reaction center of Photosystem II. Transient absorption measurements on isolated proteins reveal that photoexcitation of the high-energy core dihydrobiliverdins (DBVs, Figure 1a blue and green), is followed by transfer to the low-energy phycocyanobilin 82s (PCB82s, Figure 1a red and cyan), skipping the energetically intermediate mesobiliverdins (MBVs, Figure 1a brown and magenta) \cite{Marin2011a}. Direct down-conversion cannot be explained by inter-pigment electronic couplings \cite{Mirkovic:2007bt}, implying that the vibrational dynamics determined by local protein environments must influence the pathways of EET.

Quantifying the ultrafast vibrational dynamics of specific chromophores in LHCs remains a challenge for both theory and experiment. Using a combination of linear and non-linear spectroscopic data, one can parameterize a model that describes the local vibrational environments of each pigment \cite{Novoderezhkin:2004bg}. However, multiple distinct vibrational environments quickly lead to an intractable number of free parameters. Atomistic nuclear dynamics simulations have the potential to bypass the inverse problem \cite{Shim2012a, Viani:2014kk, Olbrich:2011dh, Rosnik:2015db} but have not previously managed to quantitatively reproduce spectroscopic signals, thus restricting their ability to inform on biological function.

For the first time, we successfully connect unique pigment-protein vibrational environments to spectroscopic signals by incorporating pigment forces calculated with ground-state density functional theory (DFT). We extract bilin reorganization energies with a wider spread and larger average value than previously expected \cite{Mirkovic:2007bt, Huo2011a, Dean2016a}, allowing for quantitative reproduction of linear spectra in the absence of free parameters. Additionally, our EET simulations yield transfer pathways and timescales in good agreement with experimental transient absorption measurements. To generalize our findings, we examine the simplest representative model system, a dimer with a resonant vibration, and define a ratio that distinguishes between the coherent and incoherent transport regimes. We find that transport between bilins in PC645 occurs incoherently, suggesting that F{\"o}rster spectral overlaps tuned by local bilin-protein environments control EET and enable direct down-conversion.

\begin{figure*}[t!]
\begin{center}
\includegraphics[width=0.88\textwidth]{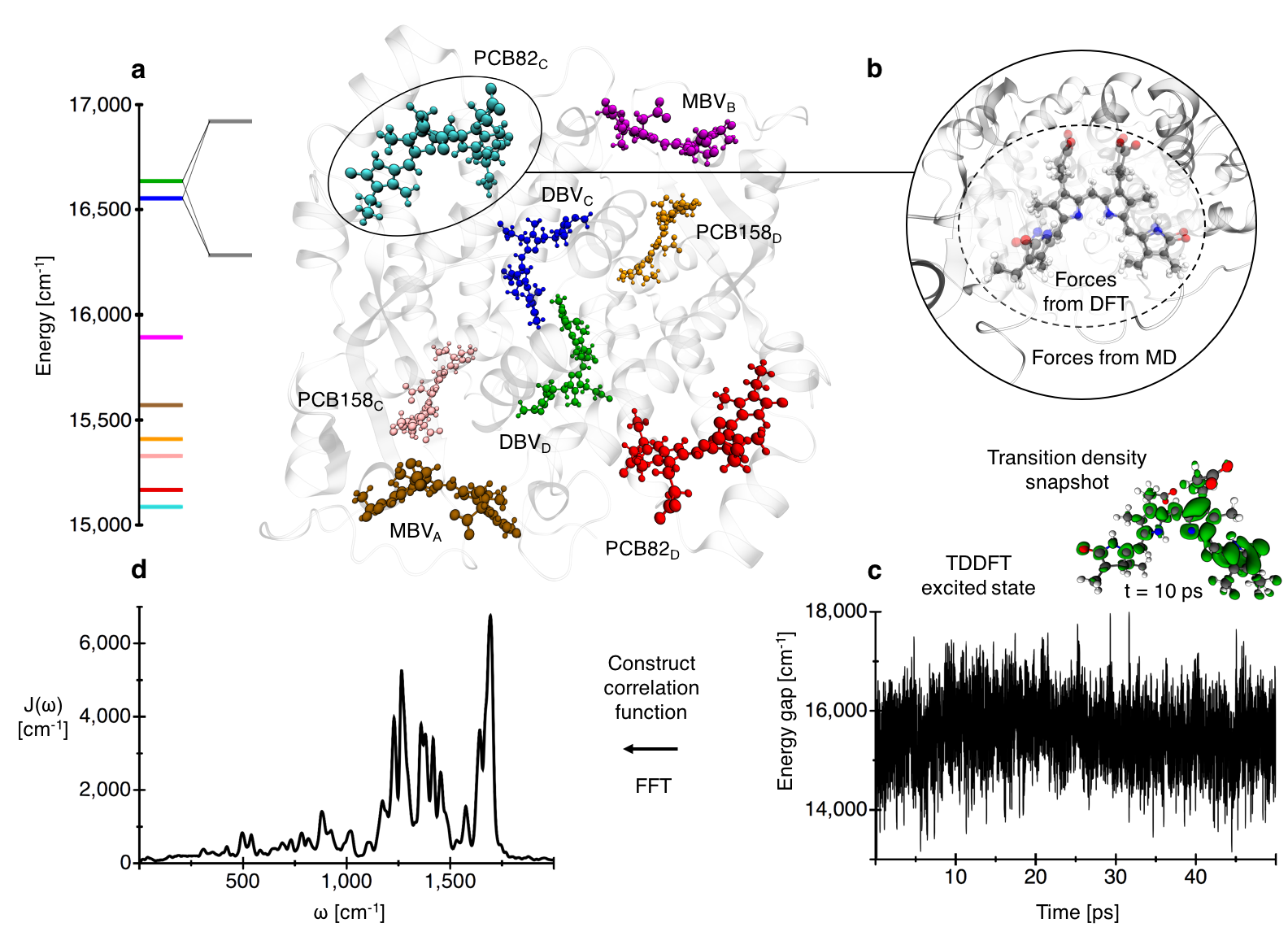}
\caption{\textbf{PC645 and computational methodology for spectral density construction. a}, Structure and reorganized energy levels of PC645, where we have defined a bilin color scheme that remains consistent throughout. Gray lines in the energy level diagram are drawn schematically to represent the exciton eigenstates of the core DBVs (blue and green) due to the significant excitonic coupling of 319.4 cm$^{-1}$, more than three times larger than any other coupling in PC645. \textbf{b}, For each bilin, we simulate 50 ps of mixed QM/MM nuclear dynamics in which that bilin is treated quantum mechanically, while the remaining protein is treated with a classical force field. \textbf{c}, We construct energy-gap trajectories by calculating full TDDFT excited states for each bilin on geometries extracted at two femtosecond intervals from our QM/MM trajectories. \textbf{d}, We construct an energy-gap correlation function for each bilin and Fourier transform to obtain eight unique bilin spectral densities.}
\end{center}
\end{figure*}

\section*{Results and discussion}
\subsection*{\textit{\textbf{Ab Initio}} Simulations of Protein Solvation Reproduce Linear Spectra}
\hspace{\parindent} To understand the role of the local pigment environments in controlling EET, we first characterize the atomistic origin of the bilin solvation dynamics. The excitation energy of a pigment bound in a protein pocket, given as the difference between the ground and first excited state energies, fluctuates due to the nuclear motions of the chromophore, the surrounding protein residues, and the proximate water. Simulating excitation energy fluctuations requires performing nuclear dynamics that sample the ground-state potential energy surface (PES) and calculating the excitation energy at regular intervals. These fluctuations can be characterized by a spectral density that describes coupling of the electronic excited state to a continuous distribution of vibrational modes \cite{Fleming:1996td}. Given the significant computational cost of obtaining nuclear forces quantum mechanically (QM), previous calculations have always used classical or semi-empirical molecular mechanics (MM) force fields to propagate nuclear dynamics for entire LHCs \cite{Shim2012a, Viani:2014kk, Olbrich:2011dh, Rosnik:2015db, Aghtar:2017dr}. However, in many cases, the MM PES sampled during nuclear dynamics differs substantially from the QM PES on which excitation energies are defined, causing the calculated spectral densities to report incorrect vibrational frequencies and coupling amplitudes \cite{Lee:2016fm}.

We construct spectral densities using nuclear dynamics calculated on an {\it{ab initio}} QM/MM PES that combines bilin nuclear gradients calculated using DFT with an MM force field to treat the surrounding protein environment (Figure 1b), thereby resolving the mismatch between nuclear dynamics trajectories and excited state calculations. Figure 1 provides an overview of our procedure for constructing spectral densities. For each bilin, we construct energy-gap trajectories (Figure 1c) by calculating excitation energies with time-dependent density functional theory (TDDFT) on 20,000 geometries extracted at two femtoseconds intervals from our 40 picosecond (ps) QM/MM production runs. Fourier transforms of the two-time correlation functions of the energy-gap trajectories define unique spectral densities that characterize individual bilin environments (Figure 1d).

Spectroscopic signals and EET dynamics depend intimately on the solvation dynamics of individual chromophores. The solvation capacity of the local vibrational environment is quantified by the total reorganization energy ($\lambda$) of the spectral density which measures the energy dissipated as the chromophore relaxes following excitation. Previous studies have assumed that all bilins have identical spectral densities and assigned a reorganization energy of either 260 cm$^{-1}$ \cite{Huo2011a}, 314 cm$^{-1}$ \cite{Dean2016a}, or 480 cm$^{-1}$ \cite{Mirkovic:2007bt}. Our spectral densities, shown in gray in the subpanels of Figure 2, reveal larger reorganization energies ( $\langle\lambda\rangle$ = 909 cm$^{-1}$) and significant variations between different bilins. The spread in our reorganization energies ($\lambda$ = 627 - 1414 cm$^{-1}$) is consistent with the presence of three chemically distinct bilins with different conjugation lengths and either one or two covalent protein linkages. The variability of the reorganization energies between chemically identical bilins bound in distinct protein environments (e.g. MBV$_{A}$ and MBV$_{B}$) demonstrates the importance of the protein scaffold in determining the local solvation dynamics. On the other hand, all eight bilin spectral densities show a peak near 1650 cm$^{-1}$ which is consistent with the assignment of a long-lived 1580 cm$^{-1}$ mode as intramolecular C=C or C=N vibrations in broad-band transient absorption measurements \cite{Dean2016a}.

Incorporating the {\it{ab initio}} QM/MM spectral densities into absorption and fluorescence simulations (solid black lines) results in excellent agreement with experimental spectra (open grey circles), as shown in Figure 2. We note that the negative features observed in simulated spectra are unphysical but do not impact the quality of the lineshapes (Figure S4). To account for both the large reorganization energies and the multiple timescales of vibrational relaxation encoded in the structure of our spectral densities, we employ numerically exact hierarchical equations of motion (HEOM) \cite{Tanimura1989a, Tanimura2012a}, implemented in QMaster \cite{Kreisbeck2014a}. HEOM has been shown to yield realistic simulations of exciton dynamics in LHCs \cite{Kreisbeck:2016cn, Strumpfer:2009jc, Ishizaki:2009tt}, but its computational complexity limits our spectral densities to including a total of 24 Drude-Lorentz peaks for full-system simulations, each representing a distributions of bath modes. We construct four classes of abridged spectral densities for each bilin starting with Class 1 (eight or nine peaks per bilin) and successively coarse-graining to incorporate fewer peaks until we reach Class 4 (two peaks per bilin). Peak parameters for all spectral densities are given in Table S1-S4. Abridged spectral densities used for spectroscopic calculations are shown with our consistent bilin color scheme in the subpanels of Figure 2. We justify our abridged spectral densities by comparing simulated monomer absorption and fluorescence spectra, as described in supplementary information section two. Supplementary information section three discusses the system Hamiltonian, {\it{ab initio}} transition dipole moments, details of the fluorescence simulations, and inhomogeneous broadening.

\begin{figure*}[t!]
\begin{center}
\includegraphics[width=0.95\textwidth]{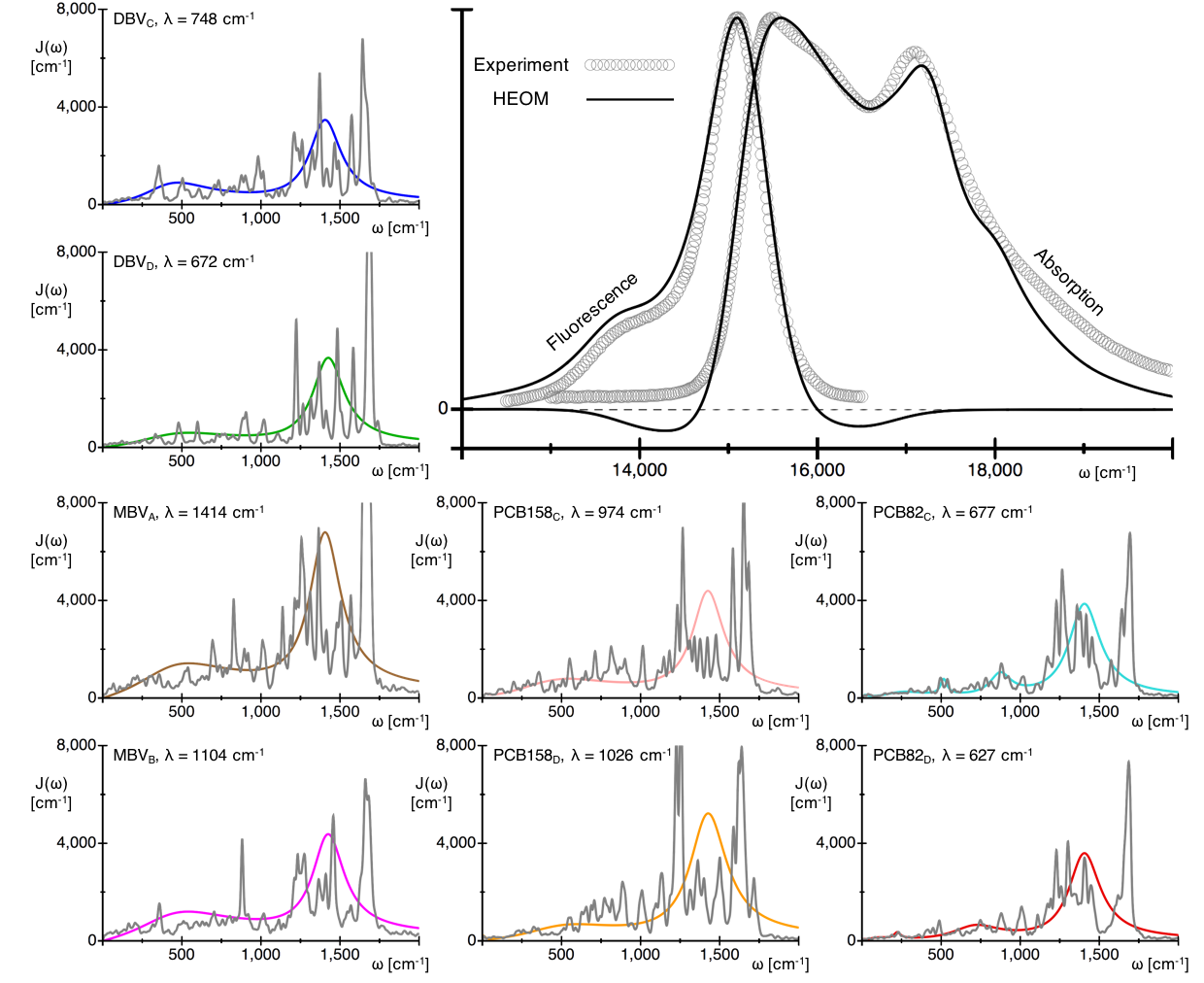}
\caption{\textbf{Comparison between absorption and fluorescence simulated with atomistic spectral densities and experimental spectra.} Unabridged (gray) and abridged (colored) bilin spectral densities are shown in the subpanels, where the latter are used for spectroscopic simulations. Each spectral density panel is labeled with the bilin name and the calculated reorganization energy $\lambda$. The main panel compares the calculated absorption and fluorescence spectra (solid black lines) with experimental spectra (open circles).}
\end{center}
\end{figure*}

\begin{figure*}[t!]
\begin{center}
\includegraphics[width=0.94\textwidth]{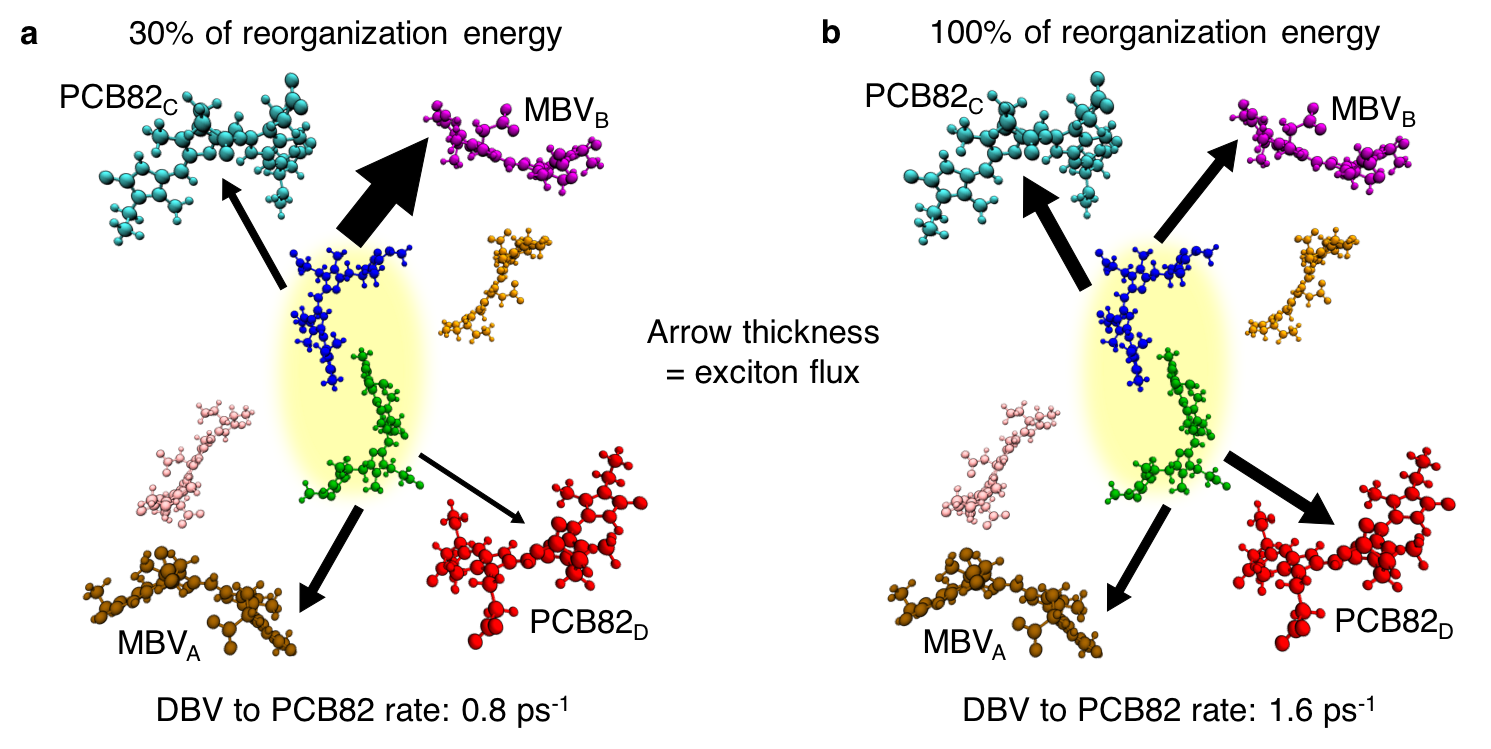}
\caption{\textbf{EET pathways and down-conversion rates simulated with HEOM.} Arrow thickness is proportional to the exciton flux leaving the DBV core integrated over one picosecond after an initial photoexcitation with (\textbf{a}) 30\% or (\textbf{b}) 100\% of calculated reorganization energy. The corresponding rates from the DBV core to both PCB82s are reported below each panel.}
\end{center}
\end{figure*}

We validate our spectral densities by comparing to two features of the experimental spectra: the Stokes shift, defined as the frequency difference between the highest energy fluorescence peak and the lowest energy absorption peak, and the overall width of the absorption spectrum. Together, these observables are quite sensitive to the distribution of reorganization energies between the different bilins of PC645. Employing the lowest energy bilin spectral density for all pigments, such as would be extracted from a fluorescence line-narrowing experiment, results in an absorption spectrum with a full-width at half maximum (FWHM) that is 25\% too small (Figure S5). The absorption spectrum contracts in this case because we underestimate the reorganization energies of the PCB158s and MBVs. If we instead use the average of the eight bilin spectral densities for all pigments, we obtain an absorption spectrum with a Stokes shift that is 55\% too large (Figure S6). The Stokes shift expands because we substantially overestimate the reorganization energy of the PCB82s that dominate the fluorescence and low-energy absorption features. Thus, 18\% error between the experimental and simulated Stokes shift and very good agreement with the overall absorption linewidth represents a strong validation of the solvation dynamics encoded in our atomistic spectral densities.

To the best of our knowledge, in the absence of free parameters, Figure 2 represents the closest reported agreement between experimental lineshapes and {\it{ab initio}} simulations of absorption and fluorescence spectra for pigments in a heterogeneous environment.

\subsection*{Protein Solvation Drives Down-Conversion}
\hspace{\parindent} The spectral density connects the atomistic dynamics of the bilin vibrational environment to both spectroscopic lineshapes and EET pertinent to light-harvesting. Having validated our atomistic QM/MM spectral densities against absorption and fluorescence spectra, we now confirm that they also describe the experimental observables associated with direct down-conversion in PC645. Experimental transient absorption measurements, combined with global kinetic analysis, indicate that initial excitation of the DBV core is followed by direct transport to the low-energy PCB82 pigments with a rate of 1.7 ps$^{-1}$ \cite{Marin2011a, Dean2016a}. Neither the rates nor selectivity of direct down-conversion have been successfully reproduced using previously estimated unstructured spectral densities \cite{Huo2011a}. Here, we show that numerically exact HEOM simulations combined with our novel spectral densities yield accurate predictions of the pathways and rates associated with down-conversion.

The large reorganization energies of our QM/MM spectral densities are essential for correctly predicting the selectivity of transport from the DBV core to the low-energy PCB82s. Simulated EET pathways following photoexcitation of the DBV core are shown in Figure 3a and 3b. Arrow thickness in Figure 3 is proportional to exciton flux, the net amount of excitation that is transferred from the DBVs to an acceptor pigment in a 1.0 ps interval following photoexcitation. We perform exciton flux calculation using Class 2 spectral densities for the MBVs and Class 4 spectral densities for the remaining pigments (supplementary information section two). In Figure 3a we rescale the spectral densities to match the average magnitude of reorganization energy used in previous simulations ($\langle\lambda\rangle$ = 260 cm$^{-1}$, labeled 30\%). In the presence of a smaller average reorganization energy, the more weakly solvated DBV core primarily transports excitation to the energetically adjacent MBV$_B$, qualitatively reproducing previous results \cite{Huo2011a}. EET simulations using the full reorganization energy (Figure 3b, labeled 100\%), however, show enhanced direct transfer to the low-energy PCB82s, in reasonable agreement with global kinetic analysis of transient absorption measurements.

In addition to influencing selectivity, the large reorganization energies also increase the rate of direct down-conversion. HEOM is a non-Markovian theory, and as a result, the rates of transport vary as a function of time in response to changes in the vibrational energy distribution. We extract a best-fit rate of down-conversion from HEOM simulations using Class 1 spectral densities with a four-site model containing only the core DBVs and the low-energy PCB82s (supplementary information section four). Class 1 spectral densities (e.g. Figure S1a) explicitly incorporate the high-frequency mode ($\sim$1650 cm$^{-1}$) previously assigned to an intramolecular bilin vibration. The rescaled spectral densities (Figure 3a, 30\%) result in a transport rate of 0.8 ps$^{-1}$, substantially slower than the experimentally observed 1.7 ps$^{-1}$. However, using the full reorganization energy of our spectral densities (Figure 3b, 100\%), we find the simulated rate of transport to be 1.6 ps$^{-1}$, within 8\% of the experimental value.

\begin{figure*}[t!]
\begin{center}
\includegraphics[width=0.9\textwidth]{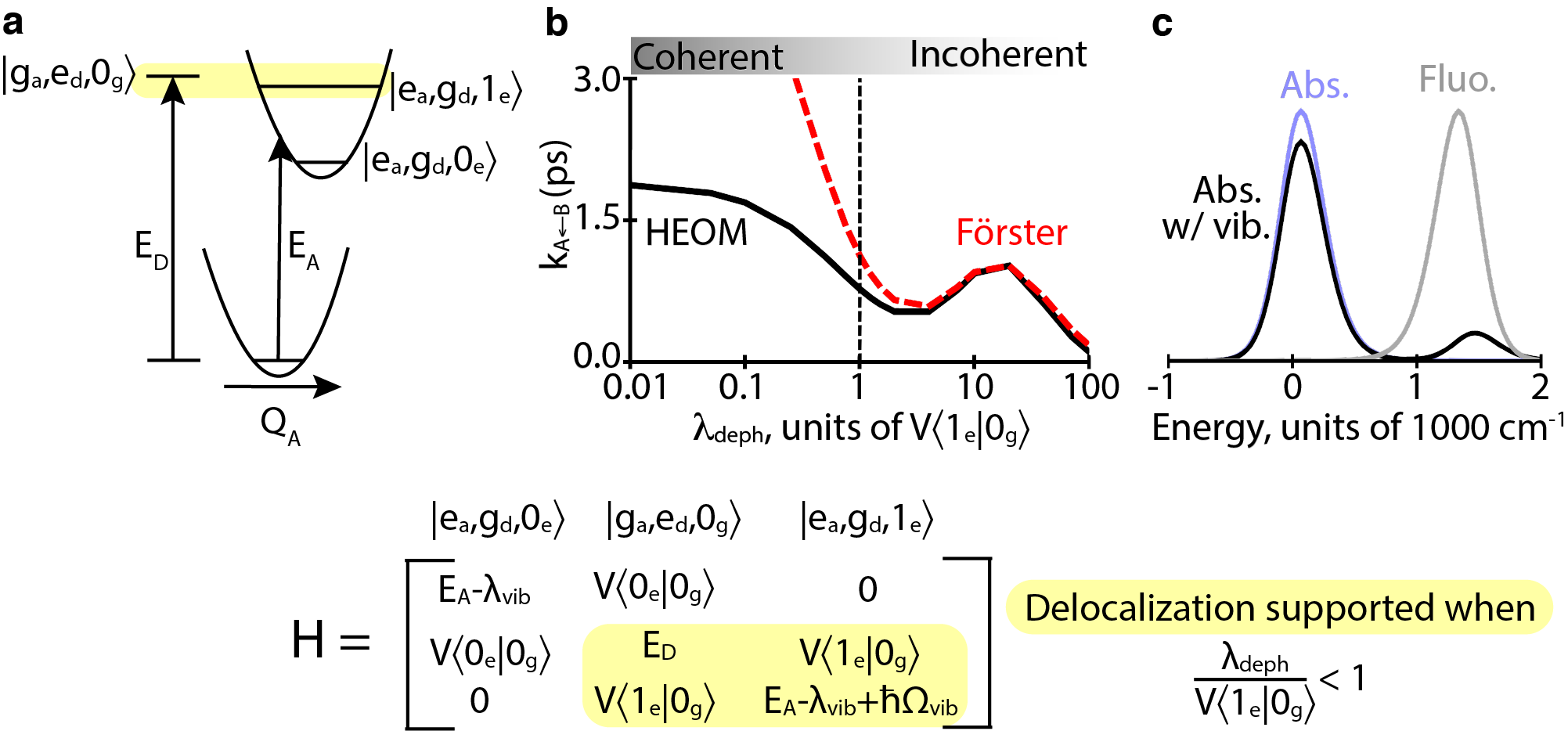}
\caption{\textbf{Regimes of coherent and incoherent vibronic transport. a}, Energy levels of a detuned model dimer with a vibration on the acceptor pigment that is nearly resonant with the site energy gap. The yellow shading represents the possible delocalization between the donor  and vibronic transition of the acceptor pigment. \textbf{b}, Rate of transport from donor  to acceptor as a function of the reorganization energy of the low-frequency background ($\lambda_{deph}$), where E$_D$ - E$_A$ = 350 cm$^{-1}$, V = 24.3 cm$^{-1}$, $\lambda_{vib}$ = 52.6 cm$^{-1}$, $\Omega_{vib}$ = 403.5 cm$^{-1}$, and $\gamma_{vib}$ = 5.2 cm$^{-1}$. HEOM rates (black) and F{\"o}rster rates (dashed red) agree in the incoherent regime, but F{\"o}rster does not accurately describe transport in the coherent regime. \textbf{c}, Mechanism of incoherent vibronic enhancement of transport, in which a nearly-resonant vibration can generate a vibronic sideband in the absorption (black) that can enhance overlap with the fluorescence (gray). Without the vibration, absorption (light blue) does not overlap with fluorescence, and thus no transport occurs. }
\end{center}
\end{figure*}

\subsection*{Down-Conversion Occurs via an Incoherent Vibronic Mechanism}
\hspace{\parindent} We have demonstrated that HEOM simulations using our atomistic spectral densities reproduce both the linear spectra and direct down-conversion of PC645, but what are the mechanistic principles connecting bilin vibrational environments to regulation of EET pathways? In particular, the relative importance of short-lived, low-frequency, intermolecular vibrations versus long-lived, high-frequency, intramolecular vibrations remains controversial \cite{Alan2016, Perlik:2015bc}. In this section, we assess how the mechanism of EET in PC645 arises from the interplay of inter-pigment couplings with structured spectral densities. First, we introduce the simplest representative model system that captures the essential features of vibronic transport in the presence of a long-lived high-frequency vibration and define a ratio that distinguishes between the coherent and incoherent transport regimes. Second, we demonstrate that PC645 experiences incoherent vibronic transport and that the long-lived near-resonant vibration assigned to support vibronic coherence plays no functional role in transport. Finally, we show that generalized F{\"o}rster theory explains the mechanism of transport in PC645, consistent with recent work on other PPC aggregates \cite{Novoderezhkin:2011ha, Bennett:2013gi, Raszewski:2008go, Amarnath:2016cv, Kreisbeck2014a}, and that local protein solvation enables direct down-conversion by controlling F{\"o}rster spectral overlaps.

The transition from coherent to incoherent vibronic transport is governed by the ratio of the vibronic coupling ($V_{vib}$) to the rapidly relaxing component of the pigment reorganization energy ($\lambda_{deph}$). Figure 4a depicts an energetically detuned dimer with a long-lived high-frequency vibration on the low-energy acceptor pigment. We define the basis as a direct-product of three indices: the electronic state of the acceptor ($\vert g_a \rangle$, $\vert e_a \rangle$), the electronic state of the donor ($\vert g_d \rangle$, $\vert e_d \rangle$), and the vibrational state of the acceptor on either the ground- or excited-state harmonic oscillator ($\vert 0_g \rangle$, $\vert 0_e \rangle$, $\vert 1_e \rangle$).  We note that the conclusions drawn from this model hold whether or not the vibration is placed on the acceptor, donor or both \cite{Forster1965}. Delocalization between the nearly degenerate states ($\vert$g$_a$, e$_d$, 0$_g\rangle$, $\vert$e$_a$, g$_d$, 1$_e\rangle$) is sustained by $V_{vib}$ (eq. \ref{eq_Vvib}), the product of the electronic coupling (V) and the Franck-Condon factor\cite{Forster1965}.  
\begin{equation}
\label{eq_Vvib}
V_{vib} = V \cdot \langle\textrm{1}_{\textrm{e}}\vert\textrm{0}_{\textrm{g}}\rangle \sim V \cdot\sqrt{\frac{\lambda_{vib}}{\Omega_{vib}}}
\end{equation}
Simultaneously, the inertial relaxation of low-frequency intermolecular vibrations on timescales faster than transport ($\lambda_{deph}$) drives excitations to localize on individual pigments \cite{Fleming:1996tda,Jordanides:1999eo,Ishizaki:2010fx}. In order for vibronic transport to occur coherently (i.e. aided by delocalization), the vibronic coupling should be larger than the rapidly reorganizing component of solvation (i.e. $\lambda_{deph}~\ll$~V$\cdot \langle$1$_e\vert$0$_g\rangle$). In contrast, as $\lambda_{deph}$ becomes greater than vibronic coupling, the excitation localizes and transport can be described by incoherent hopping. In the incoherent limit, assuming vibrational relaxation is not too slow, F{\"o}rster theory captures the rate of excitons hopping between pigments as given by eq. \ref{eq_Forster},

\begin{equation}
\label{eq_Forster}
K_{a\leftarrow d}=\frac{|V_{d,a}|^{2}}{\pi \hbar^{2}}\textrm{Re}[\int_0^{\infty} A_{a}(t) F_{d}^{*}(t) dt]
\end{equation}
where the V is the electronic coupling, A$_{a}$(t) is the absorption lineshape of the acceptor and F$_{d}$(t) is the fluorescence lineshape of the donor. In the coherent regime, where $\lambda_{deph}$~$\ll$~V$\langle$1$_e\vert$0$_g\rangle$, F{\"o}rster theory (dashed red) overestimates the rate of transport compared to HEOM (solid black) because it does not account for the possibility of back-transfer between the nearly degenerate states (Figure 4b). In the incoherent regime, where $\lambda_{deph}$~$\gg$~V$\langle$1$_e\vert$0$_g\rangle$, HEOM and F{\"o}rster theory predict equivalent transport rates. The F{\"o}rster rate between the donor and acceptor pigments is determined by Fermi’s golden rule and requires dissipating the energy difference into the vibrational environment. Thus the presence of a resonant, intramolecular vibration can enhance incoherent transport by providing a channel for dissipating the excess energy, as seen by the presence of a vibronic sideband that increases the absorption/fluorescence overlap (Figure 4c).

\begin{figure*}[t!]
\begin{center}
\includegraphics[width=0.95\textwidth]{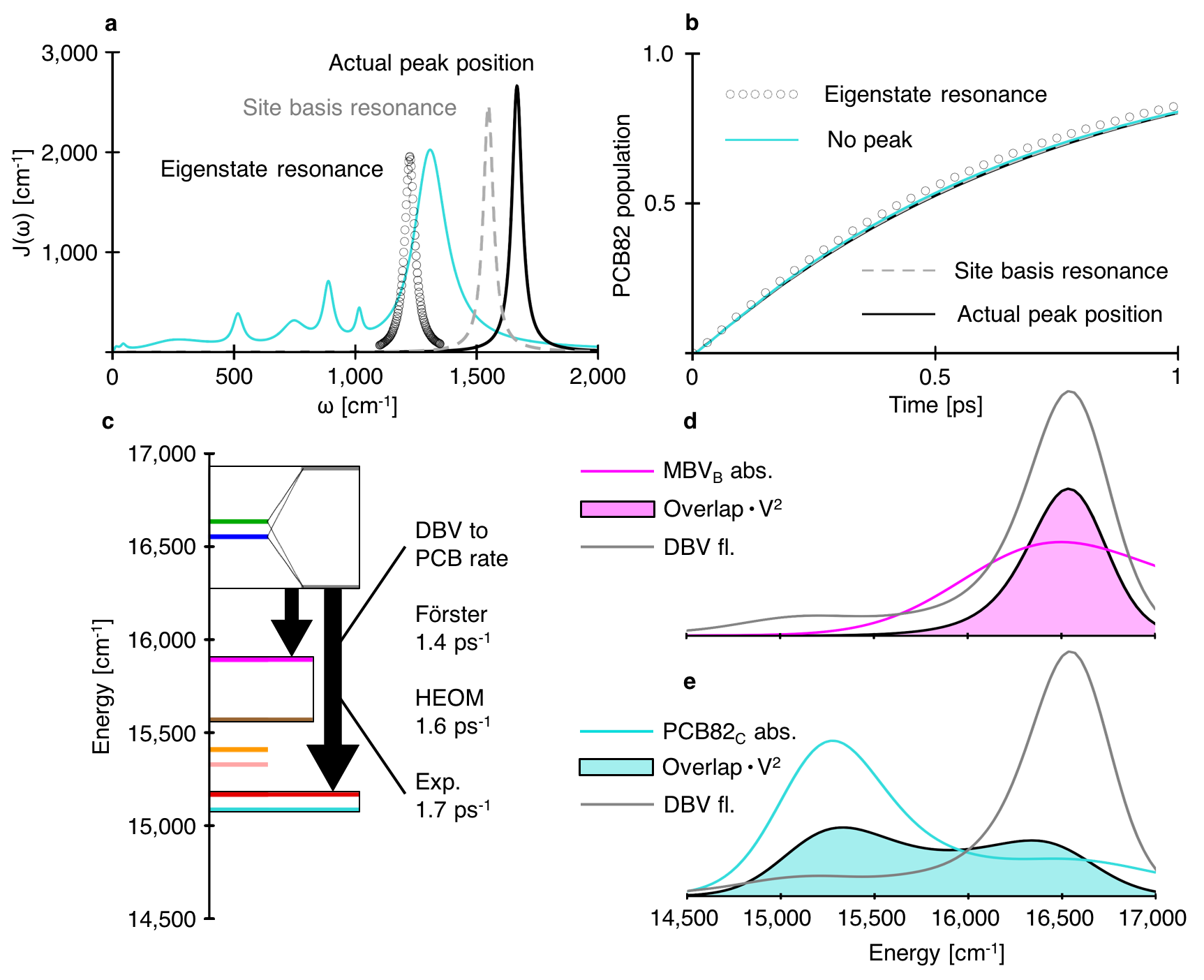}
\caption{\textbf{Direct down-conversion in PC645 is driven by incoherent vibronic transport. a}, Four different PCB82$_C$ spectral densities that we use to examine the impact of the high-frequency vibration on direct down-conversion. While cyan represents the case where the high-frequency vibration has been removed entirely, it is also implicitly added to the following three cases in which the high-frequency vibration is shown by itself for clarity: the vibration in its original position in our QM/MM spectral densities (black), the vibration shifted to be in resonance with the DBV$_D$ - PCB82$_C$ site energy difference (1550 cm$^{-1}$, dashed gray), the vibration shifted to be in resonance with the low-energy DBV exciton - PCB82$_C$ energy difference (1220 cm$^{-1}$, black circles). Note that all four sites present in our population dynamics simulations use their individual spectral densities, but the changes to the high-frequency peak are equivalently applied to the other three sites. \textbf{b}, Population dynamics of a four-site system containing the DBVs and PCB82s. Lines representing the sum PCB82 population are labeled by their representation in panel a. \textbf{c}, An energetic depiction of down-conversion where site energies are fully reorganized and arrows represent the combined flux from the DBV core to the MBVs or PCB82s. The flux arrow from DBVs to PCB82s is labeled with the effective transport rate obtained from theory and experiment. Note that excitons are drawn schematically to represent DBV delocalization. \textbf{d}, Kubo monomer absorption for MBV$_B$ (magenta), Kubo DBV core fluorescence (gray), and absorption / fluorescence overlap weighted by the coupling to the lowest energy DBV exciton (outlined magenta). \textbf{e}, Kubo monomer absorption for PCB82$_C$ (cyan), Kubo DBV core fluorescence (gray), and absorption / fluorescence overlap weighted by the coupling to the lowest energy DBV exciton (outlined cyan).}
\end{center}
\end{figure*}

We find that direct down-conversion in PC645 occurs in the incoherent regime of vibronic transport. Limited delocalization between the DBVs due to their strong electronic coupling and the many timescales of vibrational relaxation complicates the assignment of both $\lambda_{deph}$ ($\gg$~250~cm$^{-1}$) and the electronic coupling between the DBVs and PCB82s ($\sim$50 cm$^{-1}$), as described in supplementary information section four. Additionally, the high-frequency vibration on each of the DBVs and PCB82s has a Franck-Condon factor of $\sim$0.25, supporting a vibronic coupling of at most 12 cm$^{-1}$. Thus, we find that the ratio of $\lambda_{deph}$ to the vibronic coupling in PC645 is at least 20, indicating the dominance of incoherent vibronic transport. 

The presence of long-lived oscillations in nonlinear spectroscopic measurements led to the suggestion of functionally relevant vibronic coherence in PC645 \cite{Dean2016a}. We directly probe this hypothesis by manipulating the high-frequency vibration previously assigned to support vibronic signatures and demonstrate that the high-frequency mode plays no functional role in transport. Coherent vibronic transport is known to exhibit a sharp resonance condition as a function of the vibrational frequency \cite{Novoderezhkin:2017jt, Dean2016a}. The resonance condition arises because vibronic delocalization between the donor and the vibrationally excited acceptor only occurs when the energy gap is smaller than, or the same order of magnitude as, the vibronic coupling (i.e. $V_{vib} \sim 12$ cm$^{-1}$). In the case of PC645, partial delocalization in the DBVs means that there are two possible resonance conditions to consider depending on whether the DBV core eigen- or pigment-states act as the donor. To explore the possible resonance conditions, we examine population dynamics calculated in a four-site model containing only the DBVs and PCB82s. We use Class 1 spectral densities with four possible positions of the high-frequency mode (Figure 5a): the original positions ($\sim$1650 cm$^{-1}$), the DBV$_D$-PCB82$_C$ site energy difference (1550 cm$^{-1}$), the low-energy DBV exciton-PCB82$_C$ energy difference (1220 cm$^{-1}$), and the complete removal of the high-frequency peak. We observe minimal differences in the resulting population dynamics (Figure 5b). An additional exhaustive scan of the peak position between 1100 cm$^{-1}$ and 1800 cm$^{-1}$ shows no greater variations in population dynamics. We have therefore demonstrated the absence of a sharp resonance condition, further supporting our assignment of incoherent vibronic transport.

Consistent with the incoherent transport regime, F{\"o}rster theory captures the dominant contributions to the rate of down-conversion in PC645. In order to describe incoherent transport in the presence of strong coupling between DBVs, we use a generalization of F{\"o}rster theory \cite{Sumi:1999dt} that treats excitations within the DBV core as delocalized but assumes transport out of the core can be described as an incoherent hop. Due to rapid exciton relaxation within the DBV core, the rate of transport out of the core is determined by the overlap between the lowest-energy DBV exciton fluorescence and the absorption spectra of the remaining pigments. We simulate absorption spectra for the non-DBV pigments using Kubo lineshapes, which are exact for local excitations. We simulate the low-energy DBV exciton fluorescence using an extension of Kubo theory \cite{Bennett:2013gi,Sumi:1999dt,Novoderezhkin:2011ha}, which neglects the localization of excitons resulting from vibrational fluctuations (Figure S10). Generalized F{\"o}rster theory predicts a rate of 1.4 ps$^{-1}$, within 15\% of the HEOM result (Figure 5c).

Using generalized F{\"o}rster theory, we can explain the relative mechanistic role of low-frequency, intermolecular vibrations and high-frequency, intramolecular vibrations in controlling direct down-conversion in PC645. To understand the impact of bilin vibrational environments on transport rates and pathways, we examine the absorption/fluorescence overlap between the DBV core (grey lines) and either PCB82$_C$ (cyan, Figure 5d) or MBV$_B$ (magenta, Figure 5e). The Stokes shift of the DBV core ($\sim$280 cm$^{-1}$) enhances the energetic overlap with the lower-energy pigments and dramatically increases the overall rates of transport. Despite almost perfect alignment of the DBV fluorescence with the MBV absorption spectra, excitation is preferentially transported to PCB82s due to the presence of broad vibronic sidebands. In contrast to the narrow resonance condition associated with coherent vibronic transport (width on the order of $\vert$V$\langle$1$_e\vert$0$_g\rangle\vert$), the existence of wide absorption and fluorescence features allows for incoherent vibronic enhancement with a broad resonance condition (width on the order of the homogeneous linewidth). Finally, we note that given the importance of the Stokes shift in determining the rate and pathways of excitation transport, the demonstration of modified local solvation dynamics for chemically identical bilins within PC645 represents a potential design principle for nanoscale control of EET.

\section*{Conclusion}
\hspace{\parindent} The inability to connect pigment-protein solvation dynamics and regimes of EET has previously concealed the underlying design principles at play in photosynthetic LHCs. We have demonstrated how specific features of the pigment vibrational environments in PC645 exert nanoscale control of EET pathways and enable direct down-conversion. We accessed the atomistic origin of local solvation for each bilin using {\it{ab initio}} QM/MM nuclear dynamics. Our simulations identified disparities between bilin vibrational environments, which had not been sucessfully extracted from experimental measurements, yielding excellent agreement with absorption and fluorescence spectra. In contrast to previous interpretations, we have identified that down-conversion proceeds via an incoherent vibronic transport mechanism where: 
\begin{itemize}
\item excitations are localized on individual bilins (except for the DBV core) and transport occurs via incoherent hops,
\item direct down-conversion is enhanced by large reorganization energies and the presence of a wide collection of high-frequency vibrations that induce broad vibronic sidebands. 
\end{itemize}
Thus, our findings indicate that coherent vibronic transport is not relevant for biological light-harvesting in cryptophyte algal antenna complexes.

The incoherent vibronic mechanism assigned here to PC645 is far more robust to imperfections than its coherent counterpart and could act as a blueprint for the design of artificial excitonic materials. Recent advances in biomimetic light-harvesting technologies have demonstrated novel architectures, e.g. metal-organic frameworks (MOFs) \cite{So:2015fb} and DNA origami \cite{Hemmig:2016ge}, that can precisely place pigments within a hierarchically organized assembly. To fully realize nanoscale control of EET, we must iteratively simulate and design the local vibrational environments of pigment-scaffold architectures. However, the complexity of our current procedure is infeasible for high-throughput application, pointing towards the need for new and more efficient computational strategies and approximations.

\section*{Methods}
\hspace{\parindent} We perform eight QM/MM nuclear dynamics simulations where the forces on one bilin are constructed from ground-state DFT calculations for each trajectory. We begin each QM/MM trajectory with 10 picoseconds of equilibration followed by 40 picoseconds of production run. We construct energy gap trajectories for each bilin using time-dependent density functional theory (TDDFT) calculations on geometries sampled every two femtoseconds, for a total of 20,000 geometries per bilin. We use the energy gap trajectories to construct two-time correlation functions and then Fourier transform to obtain the spectral density for each bilin. Overall, the eight 50 picosecond QM/MM trajectories took over nine months to run, and cost more than two million CPU hours. The details of the initial geometry preparation, separation of the quantum from classical regions, excited-state energy calculations, and the construction of the spectral densities is described in supplementary information section five. 

Exciton dynamics simulations were performed with the QMaster software package \cite{Kreisbeck2014a} that provides a high-performance implementation of HEOM, which runs flexibly on both GPU and CPU architectures \cite{Kreisbeck2011a, Kreisbeck2012a, Kreisbeck2014a}. HEOM fluorescence calculations used a development version of QMaster. All HEOM results presented were run at a hierarchy depth of six except for rate calculations, which were run at a hierarchy depth of five (supplementary information section four).

\section*{Acknowledgements}
\hspace{\parindent} We acknowledge helpful conversations with Jacob Dean. We thank Thomas Markovich for help obtaining correlation functions and spectral densities. We acknowledge the Center for Excitonics, an Energy Frontier Research Center funded by the U.S. Department of Energy, Office of Science and Office of Basic Energy Sciences, under Award Number DE-SC0001088. S.M.B. acknowledges support from the United States Department of Energy through the Computational Sciences Graduate Fellowship (CSGF). D.I.G.B. and A.A.G. acknowledge the John Templeton Foundation (Grant Number 60469) and CIFAR, the Canadian Institute for Advanced Research, for support through the Bio-Inspired Solar Energy program. This research used resources of the National Energy Research Scientific Computing Center, a DOE Office of Science User Facility supported by the Office of Science of the U.S. Department of Energy under Contract No. DE-AC02-05CH11231. We thank Nvidia for support via the Harvard CUDA Center of Excellence. This research used computational time on the Odyssey cluster, supported by the FAS Division of Science, Research Computing Group at Harvard University.

\bibliography{PC645}

\begin{thebibliography}{57}
\expandafter\ifx\csname natexlab\endcsname\relax\def\natexlab#1{#1}\fi
\expandafter\ifx\csname bibnamefont\endcsname\relax
  \def\bibnamefont#1{#1}\fi
\expandafter\ifx\csname bibfnamefont\endcsname\relax
  \def\bibfnamefont#1{#1}\fi
\expandafter\ifx\csname citenamefont\endcsname\relax
  \def\citenamefont#1{#1}\fi
\expandafter\ifx\csname url\endcsname\relax
  \def\url#1{\texttt{#1}}\fi
\expandafter\ifx\csname urlprefix\endcsname\relax\def\urlprefix{URL }\fi
\providecommand{\bibinfo}[2]{#2}
\providecommand{\eprint}[2][]{\url{#2}}

\bibitem[{\citenamefont{Scholes et~al.}(2017)\citenamefont{Scholes, Fleming,
  Chen, Aspuru-Guzik, Buchleitner, Coker, Engel, van Grondelle, Ishizaki, Jonas
  et~al.}}]{Alan2016}
\bibinfo{author}{\bibfnamefont{G.~D.} \bibnamefont{Scholes}},
  \bibinfo{author}{\bibfnamefont{G.~R.} \bibnamefont{Fleming}},
  \bibinfo{author}{\bibfnamefont{L.~X.} \bibnamefont{Chen}},
  \bibinfo{author}{\bibfnamefont{A.}~\bibnamefont{Aspuru-Guzik}},
  \bibinfo{author}{\bibfnamefont{A.}~\bibnamefont{Buchleitner}},
  \bibinfo{author}{\bibfnamefont{D.~F.} \bibnamefont{Coker}},
  \bibinfo{author}{\bibfnamefont{G.~S.} \bibnamefont{Engel}},
  \bibinfo{author}{\bibfnamefont{R.}~\bibnamefont{van Grondelle}},
  \bibinfo{author}{\bibfnamefont{A.}~\bibnamefont{Ishizaki}},
  \bibinfo{author}{\bibfnamefont{D.~M.} \bibnamefont{Jonas}},
  \bibnamefont{et~al.}, \bibinfo{journal}{Nature}
  \textbf{\bibinfo{volume}{543}}, \bibinfo{pages}{647} (\bibinfo{year}{2017}).

\bibitem[{\citenamefont{Scholes et~al.}(2011)\citenamefont{Scholes, Fleming,
  Olaya-Castro, and van Grondelle}}]{Scholes:2011iq}
\bibinfo{author}{\bibfnamefont{G.~D.} \bibnamefont{Scholes}},
  \bibinfo{author}{\bibfnamefont{G.~R.} \bibnamefont{Fleming}},
  \bibinfo{author}{\bibfnamefont{A.}~\bibnamefont{Olaya-Castro}},
  \bibnamefont{and} \bibinfo{author}{\bibfnamefont{R.}~\bibnamefont{van
  Grondelle}}, \bibinfo{journal}{Nat. Chem.} \textbf{\bibinfo{volume}{3}},
  \bibinfo{pages}{763} (\bibinfo{year}{2011}).

\bibitem[{\citenamefont{Caram et~al.}(2012)\citenamefont{Caram, Lewis, Fidler,
  and Engel}}]{Caram:2012cc}
\bibinfo{author}{\bibfnamefont{J.~R.} \bibnamefont{Caram}},
  \bibinfo{author}{\bibfnamefont{N.~H.~C.} \bibnamefont{Lewis}},
  \bibinfo{author}{\bibfnamefont{A.~F.} \bibnamefont{Fidler}},
  \bibnamefont{and} \bibinfo{author}{\bibfnamefont{G.~S.} \bibnamefont{Engel}},
  \bibinfo{journal}{J. Chem. Phys.} \textbf{\bibinfo{volume}{136}},
  \bibinfo{pages}{104505} (\bibinfo{year}{2012}).

\bibitem[{\citenamefont{Kruger and van Grondelle}(2016)}]{Kruger:2016cs}
\bibinfo{author}{\bibfnamefont{T.~P.~J.} \bibnamefont{Kruger}}
  \bibnamefont{and} \bibinfo{author}{\bibfnamefont{R.}~\bibnamefont{van
  Grondelle}}, \bibinfo{journal}{Phys. B: Phys. Cond. Mat.}
  \textbf{\bibinfo{volume}{480}}, \bibinfo{pages}{7} (\bibinfo{year}{2016}).

\bibitem[{\citenamefont{Muh et~al.}(2007)\citenamefont{Muh, Madjet, Adolphs,
  Abdurahman, Rabenstein, Ishikita, Knapp, and Renger}}]{Muh:2007tb}
\bibinfo{author}{\bibfnamefont{F.}~\bibnamefont{Muh}},
  \bibinfo{author}{\bibfnamefont{M.~E.-A.} \bibnamefont{Madjet}},
  \bibinfo{author}{\bibfnamefont{J.}~\bibnamefont{Adolphs}},
  \bibinfo{author}{\bibfnamefont{A.}~\bibnamefont{Abdurahman}},
  \bibinfo{author}{\bibfnamefont{B.}~\bibnamefont{Rabenstein}},
  \bibinfo{author}{\bibfnamefont{H.}~\bibnamefont{Ishikita}},
  \bibinfo{author}{\bibfnamefont{E.-W.} \bibnamefont{Knapp}}, \bibnamefont{and}
  \bibinfo{author}{\bibfnamefont{T.}~\bibnamefont{Renger}},
  \bibinfo{journal}{Proc. Natl Acad. Sci.} \textbf{\bibinfo{volume}{104}},
  \bibinfo{pages}{16862} (\bibinfo{year}{2007}).

\bibitem[{\citenamefont{Novoderezhkin et~al.}(2017)\citenamefont{Novoderezhkin,
  Romero, Prior, and van Grondelle}}]{Novoderezhkin:2017jt}
\bibinfo{author}{\bibfnamefont{V.~I.} \bibnamefont{Novoderezhkin}},
  \bibinfo{author}{\bibfnamefont{E.}~\bibnamefont{Romero}},
  \bibinfo{author}{\bibfnamefont{J.}~\bibnamefont{Prior}}, \bibnamefont{and}
  \bibinfo{author}{\bibfnamefont{R.}~\bibnamefont{van Grondelle}},
  \bibinfo{journal}{Phys. Chem. Chem. Phys.} \textbf{\bibinfo{volume}{19}},
  \bibinfo{pages}{5195} (\bibinfo{year}{2017}).

\bibitem[{\citenamefont{Fuller et~al.}(2014)\citenamefont{Fuller, Pan,
  Gelzinis, Butkus, Senlik, Wilcox, Yocum, Valkunas, Abramavicius, and
  Ogilvie}}]{Fuller:2014iz}
\bibinfo{author}{\bibfnamefont{F.~D.} \bibnamefont{Fuller}},
  \bibinfo{author}{\bibfnamefont{J.}~\bibnamefont{Pan}},
  \bibinfo{author}{\bibfnamefont{A.}~\bibnamefont{Gelzinis}},
  \bibinfo{author}{\bibfnamefont{V.}~\bibnamefont{Butkus}},
  \bibinfo{author}{\bibfnamefont{S.~S.} \bibnamefont{Senlik}},
  \bibinfo{author}{\bibfnamefont{D.~E.} \bibnamefont{Wilcox}},
  \bibinfo{author}{\bibfnamefont{C.~F.} \bibnamefont{Yocum}},
  \bibinfo{author}{\bibfnamefont{L.}~\bibnamefont{Valkunas}},
  \bibinfo{author}{\bibfnamefont{D.}~\bibnamefont{Abramavicius}},
  \bibnamefont{and} \bibinfo{author}{\bibfnamefont{J.~P.}
  \bibnamefont{Ogilvie}}, \bibinfo{journal}{Nat. Chem.}
  \textbf{\bibinfo{volume}{6}}, \bibinfo{pages}{706} (\bibinfo{year}{2014}).

\bibitem[{\citenamefont{Romero et~al.}(2014)\citenamefont{Romero, Augulis,
  Novoderezhkin, Ferretti, Thieme, Zigmantas, and van
  Grondelle}}]{Romero:2014jm}
\bibinfo{author}{\bibfnamefont{E.}~\bibnamefont{Romero}},
  \bibinfo{author}{\bibfnamefont{R.}~\bibnamefont{Augulis}},
  \bibinfo{author}{\bibfnamefont{V.~I.} \bibnamefont{Novoderezhkin}},
  \bibinfo{author}{\bibfnamefont{M.}~\bibnamefont{Ferretti}},
  \bibinfo{author}{\bibfnamefont{J.}~\bibnamefont{Thieme}},
  \bibinfo{author}{\bibfnamefont{D.}~\bibnamefont{Zigmantas}},
  \bibnamefont{and} \bibinfo{author}{\bibfnamefont{R.}~\bibnamefont{van
  Grondelle}}, \bibinfo{journal}{Nat. Phys.} \textbf{\bibinfo{volume}{10}},
  \bibinfo{pages}{676} (\bibinfo{year}{2014}).

\bibitem[{\citenamefont{Dean et~al.}(2016)\citenamefont{Dean, Tihana, Toa,
  Oblinsky, and Scholes}}]{Dean2016a}
\bibinfo{author}{\bibfnamefont{J.~C.} \bibnamefont{Dean}},
  \bibinfo{author}{\bibfnamefont{M.}~\bibnamefont{Tihana}},
  \bibinfo{author}{\bibfnamefont{Z.~S.} \bibnamefont{Toa}},
  \bibinfo{author}{\bibfnamefont{D.~G.} \bibnamefont{Oblinsky}},
  \bibnamefont{and} \bibinfo{author}{\bibfnamefont{G.~D.}
  \bibnamefont{Scholes}}, \bibinfo{journal}{Chem} \textbf{\bibinfo{volume}{1}},
  \bibinfo{pages}{858} (\bibinfo{year}{2016}).

\bibitem[{\citenamefont{Christensson et~al.}(2012)\citenamefont{Christensson,
  Kauffmann, Pullerits, and Mancal}}]{Christensson:2012gp}
\bibinfo{author}{\bibfnamefont{N.}~\bibnamefont{Christensson}},
  \bibinfo{author}{\bibfnamefont{H.~F.} \bibnamefont{Kauffmann}},
  \bibinfo{author}{\bibfnamefont{T.}~\bibnamefont{Pullerits}},
  \bibnamefont{and} \bibinfo{author}{\bibfnamefont{T.}~\bibnamefont{Mancal}},
  \bibinfo{journal}{J. Phys. Chem. B} \textbf{\bibinfo{volume}{116}},
  \bibinfo{pages}{7449} (\bibinfo{year}{2012}).

\bibitem[{\citenamefont{Chin et~al.}(2013)\citenamefont{Chin, Prior, Rosenbach,
  Caycedo-Soler, Huelga, and Plenio}}]{Chin:2013ia}
\bibinfo{author}{\bibfnamefont{A.~W.} \bibnamefont{Chin}},
  \bibinfo{author}{\bibfnamefont{J.}~\bibnamefont{Prior}},
  \bibinfo{author}{\bibfnamefont{R.}~\bibnamefont{Rosenbach}},
  \bibinfo{author}{\bibfnamefont{F.}~\bibnamefont{Caycedo-Soler}},
  \bibinfo{author}{\bibfnamefont{S.~F.} \bibnamefont{Huelga}},
  \bibnamefont{and} \bibinfo{author}{\bibfnamefont{M.~B.}
  \bibnamefont{Plenio}}, \bibinfo{journal}{Nat. Phys.}
  \textbf{\bibinfo{volume}{9}}, \bibinfo{pages}{113} (\bibinfo{year}{2013}).

\bibitem[{\citenamefont{Kreisbeck and Kramer}(2012)}]{Kreisbeck2012a}
\bibinfo{author}{\bibfnamefont{C.}~\bibnamefont{Kreisbeck}} \bibnamefont{and}
  \bibinfo{author}{\bibfnamefont{T.}~\bibnamefont{Kramer}},
  \bibinfo{journal}{J. Phys. Chem. Lett.} \textbf{\bibinfo{volume}{3}},
  \bibinfo{pages}{2828} (\bibinfo{year}{2012}).

\bibitem[{\citenamefont{Kolli et~al.}(2012)\citenamefont{Kolli, O'Reilly,
  Scholes, and Olaya-Castro}}]{Kolli2012a}
\bibinfo{author}{\bibfnamefont{A.}~\bibnamefont{Kolli}},
  \bibinfo{author}{\bibfnamefont{E.~J.} \bibnamefont{O'Reilly}},
  \bibinfo{author}{\bibfnamefont{G.~D.} \bibnamefont{Scholes}},
  \bibnamefont{and}
  \bibinfo{author}{\bibfnamefont{A.}~\bibnamefont{Olaya-Castro}},
  \bibinfo{journal}{J. Chem. Phys.} \textbf{\bibinfo{volume}{137}},
  \bibinfo{pages}{174109} (\bibinfo{year}{2012}).

\bibitem[{\citenamefont{Perlik et~al.}(2015)\citenamefont{Perlik, Seibt,
  Cranston, Cogdell, Lincoln, Savolainen, Sanda, Mancal, and
  Hauer}}]{Perlik:2015bc}
\bibinfo{author}{\bibfnamefont{V.}~\bibnamefont{Perlik}},
  \bibinfo{author}{\bibfnamefont{J.}~\bibnamefont{Seibt}},
  \bibinfo{author}{\bibfnamefont{L.~J.} \bibnamefont{Cranston}},
  \bibinfo{author}{\bibfnamefont{R.~J.} \bibnamefont{Cogdell}},
  \bibinfo{author}{\bibfnamefont{C.~N.} \bibnamefont{Lincoln}},
  \bibinfo{author}{\bibfnamefont{J.}~\bibnamefont{Savolainen}},
  \bibinfo{author}{\bibfnamefont{F.}~\bibnamefont{Sanda}},
  \bibinfo{author}{\bibfnamefont{T.}~\bibnamefont{Mancal}}, \bibnamefont{and}
  \bibinfo{author}{\bibfnamefont{J.}~\bibnamefont{Hauer}}, \bibinfo{journal}{J.
  Chem. Phys.} \textbf{\bibinfo{volume}{142}}, \bibinfo{pages}{212434}
  (\bibinfo{year}{2015}).

\bibitem[{\citenamefont{Novelli et~al.}(2015)\citenamefont{Novelli, Nazir,
  Richards, Roozbeh, Wilk, Curmi, and Davis}}]{Novelli:2015es}
\bibinfo{author}{\bibfnamefont{F.}~\bibnamefont{Novelli}},
  \bibinfo{author}{\bibfnamefont{A.}~\bibnamefont{Nazir}},
  \bibinfo{author}{\bibfnamefont{G.~H.} \bibnamefont{Richards}},
  \bibinfo{author}{\bibfnamefont{A.}~\bibnamefont{Roozbeh}},
  \bibinfo{author}{\bibfnamefont{K.~E.} \bibnamefont{Wilk}},
  \bibinfo{author}{\bibfnamefont{P.~M.~G.} \bibnamefont{Curmi}},
  \bibnamefont{and} \bibinfo{author}{\bibfnamefont{J.~A.} \bibnamefont{Davis}},
  \bibinfo{journal}{J. Phys. Chem. Lett.} \textbf{\bibinfo{volume}{6}},
  \bibinfo{pages}{4573} (\bibinfo{year}{2015}).

\bibitem[{\citenamefont{Marin et~al.}(2011)\citenamefont{Marin, Doust, Scholes,
  Wilk, Curmi, van Stokkum, and van Grondelle}}]{Marin2011a}
\bibinfo{author}{\bibfnamefont{A.}~\bibnamefont{Marin}},
  \bibinfo{author}{\bibfnamefont{A.~B.} \bibnamefont{Doust}},
  \bibinfo{author}{\bibfnamefont{G.~D.} \bibnamefont{Scholes}},
  \bibinfo{author}{\bibfnamefont{K.~E.} \bibnamefont{Wilk}},
  \bibinfo{author}{\bibfnamefont{P.~M.} \bibnamefont{Curmi}},
  \bibinfo{author}{\bibfnamefont{I.~H.} \bibnamefont{van Stokkum}},
  \bibnamefont{and} \bibinfo{author}{\bibfnamefont{R.}~\bibnamefont{van
  Grondelle}}, \bibinfo{journal}{Biophys. J.} \textbf{\bibinfo{volume}{101}},
  \bibinfo{pages}{1004} (\bibinfo{year}{2011}).

\bibitem[{\citenamefont{Mirkovic et~al.}(2007)\citenamefont{Mirkovic, Doust,
  Kim, Wilk, Curutchet, Mennucci, Cammi, Curmi, and Scholes}}]{Mirkovic:2007bt}
\bibinfo{author}{\bibfnamefont{T.}~\bibnamefont{Mirkovic}},
  \bibinfo{author}{\bibfnamefont{A.~B.} \bibnamefont{Doust}},
  \bibinfo{author}{\bibfnamefont{J.}~\bibnamefont{Kim}},
  \bibinfo{author}{\bibfnamefont{K.~E.} \bibnamefont{Wilk}},
  \bibinfo{author}{\bibfnamefont{C.}~\bibnamefont{Curutchet}},
  \bibinfo{author}{\bibfnamefont{B.}~\bibnamefont{Mennucci}},
  \bibinfo{author}{\bibfnamefont{R.}~\bibnamefont{Cammi}},
  \bibinfo{author}{\bibfnamefont{P.~M.~G.} \bibnamefont{Curmi}},
  \bibnamefont{and} \bibinfo{author}{\bibfnamefont{G.~D.}
  \bibnamefont{Scholes}}, \bibinfo{journal}{Photochem. Photobiol. Sci.}
  (\bibinfo{year}{2007}).

\bibitem[{\citenamefont{Novoderezhkin et~al.}(2004)\citenamefont{Novoderezhkin,
  Palacios, van Amerongen, and van Grondelle}}]{Novoderezhkin:2004bg}
\bibinfo{author}{\bibfnamefont{V.~I.} \bibnamefont{Novoderezhkin}},
  \bibinfo{author}{\bibfnamefont{M.~A.} \bibnamefont{Palacios}},
  \bibinfo{author}{\bibfnamefont{H.}~\bibnamefont{van Amerongen}},
  \bibnamefont{and} \bibinfo{author}{\bibfnamefont{R.}~\bibnamefont{van
  Grondelle}}, \bibinfo{journal}{J. Phys. Chem. B}
  \textbf{\bibinfo{volume}{108}}, \bibinfo{pages}{10363}
  (\bibinfo{year}{2004}).

\bibitem[{\citenamefont{Shim et~al.}(2012)\citenamefont{Shim, Rebentrost,
  Valleau, and Aspuru-Guzik}}]{Shim2012a}
\bibinfo{author}{\bibfnamefont{S.}~\bibnamefont{Shim}},
  \bibinfo{author}{\bibfnamefont{P.}~\bibnamefont{Rebentrost}},
  \bibinfo{author}{\bibfnamefont{S.}~\bibnamefont{Valleau}}, \bibnamefont{and}
  \bibinfo{author}{\bibfnamefont{A.}~\bibnamefont{Aspuru-Guzik}},
  \bibinfo{journal}{Biophys. J.} \textbf{\bibinfo{volume}{102}},
  \bibinfo{pages}{649} (\bibinfo{year}{2012}).

\bibitem[{\citenamefont{Viani et~al.}(2014)\citenamefont{Viani, Corbella,
  Curutchet, O'Reilly, Olaya-Castro, and Mennucci}}]{Viani:2014kk}
\bibinfo{author}{\bibfnamefont{L.}~\bibnamefont{Viani}},
  \bibinfo{author}{\bibfnamefont{M.}~\bibnamefont{Corbella}},
  \bibinfo{author}{\bibfnamefont{C.}~\bibnamefont{Curutchet}},
  \bibinfo{author}{\bibfnamefont{E.~J.} \bibnamefont{O'Reilly}},
  \bibinfo{author}{\bibfnamefont{A.}~\bibnamefont{Olaya-Castro}},
  \bibnamefont{and} \bibinfo{author}{\bibfnamefont{B.}~\bibnamefont{Mennucci}},
  \bibinfo{journal}{Phys. Chem. Chem. Phys.} \textbf{\bibinfo{volume}{16}},
  \bibinfo{pages}{16302} (\bibinfo{year}{2014}).

\bibitem[{\citenamefont{Olbrich et~al.}(2011)\citenamefont{Olbrich, Strumpfer,
  Schulten, and Kleinekathofer}}]{Olbrich:2011dh}
\bibinfo{author}{\bibfnamefont{C.}~\bibnamefont{Olbrich}},
  \bibinfo{author}{\bibfnamefont{J.}~\bibnamefont{Strumpfer}},
  \bibinfo{author}{\bibfnamefont{K.}~\bibnamefont{Schulten}}, \bibnamefont{and}
  \bibinfo{author}{\bibfnamefont{U.}~\bibnamefont{Kleinekathofer}},
  \bibinfo{journal}{J. Phys. Chem. Lett.} \textbf{\bibinfo{volume}{2}},
  \bibinfo{pages}{1771} (\bibinfo{year}{2011}).

\bibitem[{\citenamefont{Rosnik and Curutchet}(2015)}]{Rosnik:2015db}
\bibinfo{author}{\bibfnamefont{A.~M.} \bibnamefont{Rosnik}} \bibnamefont{and}
  \bibinfo{author}{\bibfnamefont{C.}~\bibnamefont{Curutchet}},
  \bibinfo{journal}{J. Chem. Theory Comput.} \textbf{\bibinfo{volume}{11}},
  \bibinfo{pages}{5826} (\bibinfo{year}{2015}).

\bibitem[{\citenamefont{Pengfei and Coker}(2011)}]{Huo2011a}
\bibinfo{author}{\bibfnamefont{H.}~\bibnamefont{Pengfei}} \bibnamefont{and}
  \bibinfo{author}{\bibfnamefont{D.~F.} \bibnamefont{Coker}},
  \bibinfo{journal}{J. Phys. Chem. Lett.} \textbf{\bibinfo{volume}{2}},
  \bibinfo{pages}{825} (\bibinfo{year}{2011}).

\bibitem[{\citenamefont{Fleming and Cho}(1996{\natexlab{a}})}]{Fleming:1996td}
\bibinfo{author}{\bibfnamefont{G.~R.} \bibnamefont{Fleming}} \bibnamefont{and}
  \bibinfo{author}{\bibfnamefont{M.}~\bibnamefont{Cho}},
  \bibinfo{journal}{Annual Review of Physical Chemistry}
  (\bibinfo{year}{1996}{\natexlab{a}}).

\bibitem[{\citenamefont{Aghtar et~al.}(2017)\citenamefont{Aghtar,
  Kleinekathofer, Curutchet, and Mennucci}}]{Aghtar:2017dr}
\bibinfo{author}{\bibfnamefont{M.}~\bibnamefont{Aghtar}},
  \bibinfo{author}{\bibfnamefont{U.}~\bibnamefont{Kleinekathofer}},
  \bibinfo{author}{\bibfnamefont{C.}~\bibnamefont{Curutchet}},
  \bibnamefont{and} \bibinfo{author}{\bibfnamefont{B.}~\bibnamefont{Mennucci}},
  \bibinfo{journal}{J. Phys. Chem. B} \textbf{\bibinfo{volume}{121}},
  \bibinfo{pages}{1330} (\bibinfo{year}{2017}).

\bibitem[{\citenamefont{Lee and Coker}(2016)}]{Lee:2016fm}
\bibinfo{author}{\bibfnamefont{M.~K.} \bibnamefont{Lee}} \bibnamefont{and}
  \bibinfo{author}{\bibfnamefont{D.~F.} \bibnamefont{Coker}},
  \bibinfo{journal}{J. Phys. Chem. Lett.} \textbf{\bibinfo{volume}{7}},
  \bibinfo{pages}{3171} (\bibinfo{year}{2016}).

\bibitem[{\citenamefont{Tanimura and Kubo}(1989)}]{Tanimura1989a}
\bibinfo{author}{\bibfnamefont{Y.}~\bibnamefont{Tanimura}} \bibnamefont{and}
  \bibinfo{author}{\bibfnamefont{R.}~\bibnamefont{Kubo}}, \bibinfo{journal}{J.
  Phys. Soc. Jpn.} \textbf{\bibinfo{volume}{58}}, \bibinfo{pages}{101}
  (\bibinfo{year}{1989}).

\bibitem[{\citenamefont{Tanimura}(2012)}]{Tanimura2012a}
\bibinfo{author}{\bibfnamefont{Y.}~\bibnamefont{Tanimura}},
  \bibinfo{journal}{J. Chem. Phys.} \textbf{\bibinfo{volume}{137}},
  \bibinfo{pages}{22A550} (\bibinfo{year}{2012}).

\bibitem[{\citenamefont{Kreisbeck et~al.}(2014)\citenamefont{Kreisbeck, Kramer,
  and Aspuru-Guzik}}]{Kreisbeck2014a}
\bibinfo{author}{\bibfnamefont{C.}~\bibnamefont{Kreisbeck}},
  \bibinfo{author}{\bibfnamefont{T.}~\bibnamefont{Kramer}}, \bibnamefont{and}
  \bibinfo{author}{\bibfnamefont{A.}~\bibnamefont{Aspuru-Guzik}},
  \bibinfo{journal}{J. Chem. Theory Comput.} \textbf{\bibinfo{volume}{10}},
  \bibinfo{pages}{4045} (\bibinfo{year}{2014}).

\bibitem[{\citenamefont{Kreisbeck and Aspuru-Guzik}(2016)}]{Kreisbeck:2016cn}
\bibinfo{author}{\bibfnamefont{C.}~\bibnamefont{Kreisbeck}} \bibnamefont{and}
  \bibinfo{author}{\bibfnamefont{A.}~\bibnamefont{Aspuru-Guzik}},
  \bibinfo{journal}{Chem. Sci.} \textbf{\bibinfo{volume}{7}},
  \bibinfo{pages}{4174} (\bibinfo{year}{2016}).

\bibitem[{\citenamefont{Strumpfer and Schulten}(2009)}]{Strumpfer:2009jc}
\bibinfo{author}{\bibfnamefont{J.}~\bibnamefont{Strumpfer}} \bibnamefont{and}
  \bibinfo{author}{\bibfnamefont{K.}~\bibnamefont{Schulten}},
  \bibinfo{journal}{J. Chem. Phys.} \textbf{\bibinfo{volume}{131}},
  \bibinfo{pages}{225101} (\bibinfo{year}{2009}).

\bibitem[{\citenamefont{Ishizaki and Fleming}(2009)}]{Ishizaki:2009tt}
\bibinfo{author}{\bibfnamefont{A.}~\bibnamefont{Ishizaki}} \bibnamefont{and}
  \bibinfo{author}{\bibfnamefont{G.~R.} \bibnamefont{Fleming}},
  \bibinfo{journal}{Proc. Natl Acad. Sci.} \textbf{\bibinfo{volume}{106}},
  \bibinfo{pages}{17255} (\bibinfo{year}{2009}).

\bibitem[{\citenamefont{Novoderezhkin et~al.}(2011)\citenamefont{Novoderezhkin,
  Marin, and van Grondelle}}]{Novoderezhkin:2011ha}
\bibinfo{author}{\bibfnamefont{V.}~\bibnamefont{Novoderezhkin}},
  \bibinfo{author}{\bibfnamefont{A.}~\bibnamefont{Marin}}, \bibnamefont{and}
  \bibinfo{author}{\bibfnamefont{R.}~\bibnamefont{van Grondelle}},
  \bibinfo{journal}{Phys. Chem. Chem. Phys.} \textbf{\bibinfo{volume}{13}},
  \bibinfo{pages}{17093} (\bibinfo{year}{2011}).

\bibitem[{\citenamefont{Bennett et~al.}(2013)\citenamefont{Bennett, Amarnath,
  and Fleming}}]{Bennett:2013gi}
\bibinfo{author}{\bibfnamefont{D.~I.~G.} \bibnamefont{Bennett}},
  \bibinfo{author}{\bibfnamefont{K.}~\bibnamefont{Amarnath}}, \bibnamefont{and}
  \bibinfo{author}{\bibfnamefont{G.~R.} \bibnamefont{Fleming}},
  \bibinfo{journal}{J. Am. Chem. Soc.} \textbf{\bibinfo{volume}{135}},
  \bibinfo{pages}{9164} (\bibinfo{year}{2013}).

\bibitem[{\citenamefont{Raszewski and Renger}(2008)}]{Raszewski:2008go}
\bibinfo{author}{\bibfnamefont{G.}~\bibnamefont{Raszewski}} \bibnamefont{and}
  \bibinfo{author}{\bibfnamefont{T.}~\bibnamefont{Renger}},
  \bibinfo{journal}{J. Am. Chem. Soc.} \textbf{\bibinfo{volume}{130}},
  \bibinfo{pages}{4431} (\bibinfo{year}{2008}).

\bibitem[{\citenamefont{Amarnath et~al.}(2016)\citenamefont{Amarnath, Bennett,
  Schneider, and Fleming}}]{Amarnath:2016cv}
\bibinfo{author}{\bibfnamefont{K.}~\bibnamefont{Amarnath}},
  \bibinfo{author}{\bibfnamefont{D.~I.~G.} \bibnamefont{Bennett}},
  \bibinfo{author}{\bibfnamefont{A.~R.} \bibnamefont{Schneider}},
  \bibnamefont{and} \bibinfo{author}{\bibfnamefont{G.~R.}
  \bibnamefont{Fleming}}, \bibinfo{journal}{Proc. Natl Acad. Sci.}
  \textbf{\bibinfo{volume}{113}}, \bibinfo{pages}{1156} (\bibinfo{year}{2016}).

\bibitem[{\citenamefont{Forster}(1965)}]{Forster1965}
\bibinfo{author}{\bibfnamefont{T.}~\bibnamefont{Forster}}, in
  \emph{\bibinfo{booktitle}{Modern quantum chemistry. Istanbul lectures. Part
  III: Action of light and organic crystals}}, edited by
  \bibinfo{editor}{\bibfnamefont{O.}~\bibnamefont{Sinanoglu}}
  (\bibinfo{publisher}{New York and London: Academic Press},
  \bibinfo{year}{1965}), pp. \bibinfo{pages}{93--137}.

\bibitem[{\citenamefont{Fleming and Cho}(1996{\natexlab{b}})}]{Fleming:1996tda}
\bibinfo{author}{\bibfnamefont{G.~R.} \bibnamefont{Fleming}} \bibnamefont{and}
  \bibinfo{author}{\bibfnamefont{M.}~\bibnamefont{Cho}}, \bibinfo{journal}{Ann.
  Rev. Phys. Chem.} \textbf{\bibinfo{volume}{47}}, \bibinfo{pages}{109}
  (\bibinfo{year}{1996}{\natexlab{b}}).

\bibitem[{\citenamefont{Jordanides et~al.}(1999)\citenamefont{Jordanides, Lang,
  Song, and Fleming}}]{Jordanides:1999eo}
\bibinfo{author}{\bibfnamefont{X.~J.} \bibnamefont{Jordanides}},
  \bibinfo{author}{\bibfnamefont{M.~J.} \bibnamefont{Lang}},
  \bibinfo{author}{\bibfnamefont{X.}~\bibnamefont{Song}}, \bibnamefont{and}
  \bibinfo{author}{\bibfnamefont{G.~R.} \bibnamefont{Fleming}},
  \bibinfo{journal}{J. Phys. Chem. B} \textbf{\bibinfo{volume}{103}},
  \bibinfo{pages}{7995} (\bibinfo{year}{1999}).

\bibitem[{\citenamefont{Ishizaki et~al.}(2010)\citenamefont{Ishizaki, Calhoun,
  Schlau-Cohen, and Fleming}}]{Ishizaki:2010fx}
\bibinfo{author}{\bibfnamefont{A.}~\bibnamefont{Ishizaki}},
  \bibinfo{author}{\bibfnamefont{T.~R.} \bibnamefont{Calhoun}},
  \bibinfo{author}{\bibfnamefont{G.~S.} \bibnamefont{Schlau-Cohen}},
  \bibnamefont{and} \bibinfo{author}{\bibfnamefont{G.~R.}
  \bibnamefont{Fleming}}, \bibinfo{journal}{Phys. Chem. Chem. Phys.}
  \textbf{\bibinfo{volume}{12}}, \bibinfo{pages}{7319} (\bibinfo{year}{2010}).

\bibitem[{\citenamefont{Sumi}(1999)}]{Sumi:1999dt}
\bibinfo{author}{\bibfnamefont{H.}~\bibnamefont{Sumi}}, \bibinfo{journal}{J.
  Phys. Chem. B} \textbf{\bibinfo{volume}{103}}, \bibinfo{pages}{252}
  (\bibinfo{year}{1999}).

\bibitem[{\citenamefont{So et~al.}(2015)\citenamefont{So, Wiederrecht,
  Mondloch, Hupp, and Farha}}]{So:2015fb}
\bibinfo{author}{\bibfnamefont{M.~C.} \bibnamefont{So}},
  \bibinfo{author}{\bibfnamefont{G.~P.} \bibnamefont{Wiederrecht}},
  \bibinfo{author}{\bibfnamefont{J.~E.} \bibnamefont{Mondloch}},
  \bibinfo{author}{\bibfnamefont{J.~T.} \bibnamefont{Hupp}}, \bibnamefont{and}
  \bibinfo{author}{\bibfnamefont{O.~K.} \bibnamefont{Farha}},
  \bibinfo{journal}{Chem. Commun.} \textbf{\bibinfo{volume}{51}},
  \bibinfo{pages}{3501} (\bibinfo{year}{2015}).

\bibitem[{\citenamefont{Hemmig et~al.}(2016)\citenamefont{Hemmig, Creatore,
  Wunsch, Hecker, Mair, Parker, Emmott, Tinnefeld, Keyser, and
  Chin}}]{Hemmig:2016ge}
\bibinfo{author}{\bibfnamefont{E.~A.} \bibnamefont{Hemmig}},
  \bibinfo{author}{\bibfnamefont{C.}~\bibnamefont{Creatore}},
  \bibinfo{author}{\bibfnamefont{B.}~\bibnamefont{Wunsch}},
  \bibinfo{author}{\bibfnamefont{L.}~\bibnamefont{Hecker}},
  \bibinfo{author}{\bibfnamefont{P.}~\bibnamefont{Mair}},
  \bibinfo{author}{\bibfnamefont{M.~A.} \bibnamefont{Parker}},
  \bibinfo{author}{\bibfnamefont{S.}~\bibnamefont{Emmott}},
  \bibinfo{author}{\bibfnamefont{P.}~\bibnamefont{Tinnefeld}},
  \bibinfo{author}{\bibfnamefont{U.~F.} \bibnamefont{Keyser}},
  \bibnamefont{and} \bibinfo{author}{\bibfnamefont{A.~W.} \bibnamefont{Chin}},
  \bibinfo{journal}{Nano Lett.} \textbf{\bibinfo{volume}{16}},
  \bibinfo{pages}{2369} (\bibinfo{year}{2016}).

\bibitem[{\citenamefont{Kreisbeck et~al.}(2011)\citenamefont{Kreisbeck, Kramer,
  Rodriguez, and Hein}}]{Kreisbeck2011a}
\bibinfo{author}{\bibfnamefont{C.}~\bibnamefont{Kreisbeck}},
  \bibinfo{author}{\bibfnamefont{T.}~\bibnamefont{Kramer}},
  \bibinfo{author}{\bibfnamefont{M.}~\bibnamefont{Rodriguez}},
  \bibnamefont{and} \bibinfo{author}{\bibfnamefont{B.}~\bibnamefont{Hein}},
  \bibinfo{journal}{J. Chem. Theory Comput.} \textbf{\bibinfo{volume}{7}},
  \bibinfo{pages}{2166} (\bibinfo{year}{2011}).

\bibitem[{\citenamefont{Dijkstra et~al.}(2015)\citenamefont{Dijkstra, Wang,
  Cao, and Fleming}}]{Dijkstra:2015er}
\bibinfo{author}{\bibfnamefont{A.~G.} \bibnamefont{Dijkstra}},
  \bibinfo{author}{\bibfnamefont{C.}~\bibnamefont{Wang}},
  \bibinfo{author}{\bibfnamefont{J.}~\bibnamefont{Cao}}, \bibnamefont{and}
  \bibinfo{author}{\bibfnamefont{G.~R.} \bibnamefont{Fleming}},
  \bibinfo{journal}{J. Phys. Chem. Lett.} \textbf{\bibinfo{volume}{6}},
  \bibinfo{pages}{627} (\bibinfo{year}{2015}).

\bibitem[{\citenamefont{Harrop et~al.}(2014)\citenamefont{Harrop, Wilk,
  Dinshaw, Collini, Mirkovic, Teng, Oblinsky, Green, Hoef-Emden, Hiller
  et~al.}}]{Harrop:2014io}
\bibinfo{author}{\bibfnamefont{S.~J.} \bibnamefont{Harrop}},
  \bibinfo{author}{\bibfnamefont{K.~E.} \bibnamefont{Wilk}},
  \bibinfo{author}{\bibfnamefont{R.}~\bibnamefont{Dinshaw}},
  \bibinfo{author}{\bibfnamefont{E.}~\bibnamefont{Collini}},
  \bibinfo{author}{\bibfnamefont{T.}~\bibnamefont{Mirkovic}},
  \bibinfo{author}{\bibfnamefont{C.~Y.} \bibnamefont{Teng}},
  \bibinfo{author}{\bibfnamefont{D.~G.} \bibnamefont{Oblinsky}},
  \bibinfo{author}{\bibfnamefont{B.~R.} \bibnamefont{Green}},
  \bibinfo{author}{\bibfnamefont{K.}~\bibnamefont{Hoef-Emden}},
  \bibinfo{author}{\bibfnamefont{R.~G.} \bibnamefont{Hiller}},
  \bibnamefont{et~al.}, \bibinfo{journal}{Proc. Natl Acad. Sci.}
  \textbf{\bibinfo{volume}{111}}, \bibinfo{pages}{E2666}
  (\bibinfo{year}{2014}).

\bibitem[{\citenamefont{D.A.~Case and Kollman}(2016)}]{AMBER2016}
\bibinfo{author}{\bibfnamefont{R.~B.} \bibnamefont{D.A.~Case}}
  \bibnamefont{and} \bibinfo{author}{\bibfnamefont{P.}~\bibnamefont{Kollman}}
  (\bibinfo{year}{2016}).

\bibitem[{\citenamefont{Valiev et~al.}(2010)\citenamefont{Valiev, Bylaska,
  Govind, and Kowalski}}]{Valiev:2010bb}
\bibinfo{author}{\bibfnamefont{M.}~\bibnamefont{Valiev}},
  \bibinfo{author}{\bibfnamefont{E.~J.} \bibnamefont{Bylaska}},
  \bibinfo{author}{\bibfnamefont{N.}~\bibnamefont{Govind}}, \bibnamefont{and}
  \bibinfo{author}{\bibfnamefont{K.}~\bibnamefont{Kowalski}},
  \bibinfo{journal}{Comput. Phys.} \textbf{\bibinfo{volume}{181}},
  \bibinfo{pages}{1477} (\bibinfo{year}{2010}).

\bibitem[{\citenamefont{Singh and Kollman}(1986)}]{Singh:1986cu}
\bibinfo{author}{\bibfnamefont{U.~C.} \bibnamefont{Singh}} \bibnamefont{and}
  \bibinfo{author}{\bibfnamefont{P.~A.} \bibnamefont{Kollman}},
  \bibinfo{journal}{J. Comput. Chem.} \textbf{\bibinfo{volume}{7}},
  \bibinfo{pages}{718} (\bibinfo{year}{1986}).

\bibitem[{\citenamefont{Lee et~al.}(1988)\citenamefont{Lee, Yang, and
  Parr}}]{Lee:1988fm}
\bibinfo{author}{\bibfnamefont{C.}~\bibnamefont{Lee}},
  \bibinfo{author}{\bibfnamefont{W.}~\bibnamefont{Yang}}, \bibnamefont{and}
  \bibinfo{author}{\bibfnamefont{R.~G.} \bibnamefont{Parr}},
  \bibinfo{journal}{Phys. Rev. B} \textbf{\bibinfo{volume}{37}},
  \bibinfo{pages}{785} (\bibinfo{year}{1988}).

\bibitem[{\citenamefont{Becke}(1993)}]{Becke:1993is}
\bibinfo{author}{\bibfnamefont{A.~D.} \bibnamefont{Becke}},
  \bibinfo{journal}{J. Chem. Phys.} \textbf{\bibinfo{volume}{98}},
  \bibinfo{pages}{5648} (\bibinfo{year}{1993}).

\bibitem[{\citenamefont{Stephens et~al.}(1994)\citenamefont{Stephens, Devlin,
  and Chabalowski}}]{Stephens:1994jt}
\bibinfo{author}{\bibfnamefont{P.~J.} \bibnamefont{Stephens}},
  \bibinfo{author}{\bibfnamefont{F.~J.} \bibnamefont{Devlin}},
  \bibnamefont{and} \bibinfo{author}{\bibfnamefont{C.~F.}
  \bibnamefont{Chabalowski}}, \bibinfo{journal}{J. Phys. Chem.}
  \textbf{\bibinfo{volume}{98}} (\bibinfo{year}{1994}).

\bibitem[{\citenamefont{Berendsen et~al.}(1984)\citenamefont{Berendsen, Postma,
  van Gunsteren, DiNola, and Haak}}]{Berendsen:1984fm}
\bibinfo{author}{\bibfnamefont{H.~J.~C.} \bibnamefont{Berendsen}},
  \bibinfo{author}{\bibfnamefont{J.~P.~M.} \bibnamefont{Postma}},
  \bibinfo{author}{\bibfnamefont{W.~F.} \bibnamefont{van Gunsteren}},
  \bibinfo{author}{\bibfnamefont{A.}~\bibnamefont{DiNola}}, \bibnamefont{and}
  \bibinfo{author}{\bibfnamefont{J.~R.} \bibnamefont{Haak}},
  \bibinfo{journal}{J. Chem. Phys.} \textbf{\bibinfo{volume}{81}},
  \bibinfo{pages}{3684} (\bibinfo{year}{1984}).

\bibitem[{\citenamefont{Morishita}(2000)}]{Morishita:2000jz}
\bibinfo{author}{\bibfnamefont{T.}~\bibnamefont{Morishita}},
  \bibinfo{journal}{J. Chem. Phys.} \textbf{\bibinfo{volume}{113}},
  \bibinfo{pages}{2976} (\bibinfo{year}{2000}).

\bibitem[{\citenamefont{Sinnokrot and Sherrill}(2001)}]{Sinnokrot:2001is}
\bibinfo{author}{\bibfnamefont{M.~O.} \bibnamefont{Sinnokrot}}
  \bibnamefont{and} \bibinfo{author}{\bibfnamefont{C.~D.}
  \bibnamefont{Sherrill}}, \bibinfo{journal}{J. Chem. Phys.}
  \textbf{\bibinfo{volume}{115}}, \bibinfo{pages}{2439} (\bibinfo{year}{2001}).

\bibitem[{\citenamefont{Falzon et~al.}(2005)\citenamefont{Falzon, Chong, and
  Wang}}]{Falzon:2005dk}
\bibinfo{author}{\bibfnamefont{C.~T.} \bibnamefont{Falzon}},
  \bibinfo{author}{\bibfnamefont{D.~P.} \bibnamefont{Chong}}, \bibnamefont{and}
  \bibinfo{author}{\bibfnamefont{F.}~\bibnamefont{Wang}}, \bibinfo{journal}{J.
  Comput. Chem.} \textbf{\bibinfo{volume}{27}}, \bibinfo{pages}{163}
  (\bibinfo{year}{2005}).

\bibitem[{\citenamefont{Valleau et~al.}(2012)\citenamefont{Valleau, Eisfeld,
  and Aspuru-Guzik}}]{Valleau:2012ig}
\bibinfo{author}{\bibfnamefont{S.}~\bibnamefont{Valleau}},
  \bibinfo{author}{\bibfnamefont{A.}~\bibnamefont{Eisfeld}}, \bibnamefont{and}
  \bibinfo{author}{\bibfnamefont{A.}~\bibnamefont{Aspuru-Guzik}},
  \bibinfo{journal}{J. Chem. Phys.} \textbf{\bibinfo{volume}{137}},
  \bibinfo{pages}{224103} (\bibinfo{year}{2012}).

\end{thebibliography}
\newpage
%\phantom{a}
\newpage

\section{Supplementary Information}
\renewcommand{\thefigure}{S\arabic{figure}}
\renewcommand{\thetable}{S\arabic{table}}

\setcounter{figure}{0}
\setcounter{equation}{0}
\newenvironment{myequation}{%
\addtocounter{equation}{-1}
\refstepcounter{defcounter}
\renewcommand\theequation{S\thedefcounter}
\begin{equation}}
{\end{equation}}

\maketitle
%\appendix
\section*{System Hamiltonian}
\addcontentsline{toc}{section}{System Hamiltonian}
\hspace{\parindent} We describe energy transfer using a Frenkel exciton Hamiltonian and assume that only one of the pigments is excited at a time. The Hamiltonian of the single exciton manifold reads
\begin{myequation}\label{eq:Hex}
H_{\rm ex}=\sum_{m=1}^N \epsilon_m^0 |m\rangle\langle m| + \sum_{m>n} J_{mn}(|m\rangle\langle n|+|n\rangle\langle m|).
\end{myequation}
Here $|m\rangle$ denotes the state in which pigment~$m$ is excited while the other pigments remain in the electronic ground state. 
$J_{mn}$ denotes the electronic coupling between the excited states on pigments~$m$ and~$n$.

The pigments are coupled to the protein environment modeled by a set of independent harmonic oscillators 
\begin{myequation}
\mathcal{H}_{\rm phon}=\sum_{m,i}\hbar\omega_i b_{i,m}^\dag b_{i,m},
\end{myequation}
and we assume a linear coupling of the exciton system to the vibrations 
\begin{myequation}
\mathcal{H}_{\rm ex-phon}=\sum_m |m\rangle\langle m|\,\sum_i\hbar\omega_{i,m}d_{i,m}(b_{i,m}+b_{i,m}^\dag).
\end{myequation}
The reorganization energy, $\lambda_m=\sum_i\hbar\omega_{i,m}d_{i,m}^2/2$, is added to the exciton energies in eqn~\eqref{eq:Hex}, $\varepsilon_m=\varepsilon_m^0+\lambda_m$. 
The phonon mode dependent coupling strength is captured by the spectral density
\begin{myequation}\label{eq:SpecDens}
 J_m(\omega)=\pi\sum_\xi \hbar^2\omega_{\xi,m}^2 d_{\xi,m}^2\delta(\omega-\omega_{\xi,m})
\end{myequation}
which can also be defined as a sum of Drude-Lorentz peaks of form

\begin{myequation}
J(\omega) =  \frac{\beta \hbar }{\sqrt{\pi}} \sum^{N_\textrm{peaks}}_{i=1,s=\pm} \frac{\lambda_{i} \gamma_i \omega}{2}  \frac{ 1}{\gamma_i^2 + \left( \omega + s\, \Omega_i \right)^2 }.
\end{myequation}
where $\lambda_i$ is the peak reorganization energy, $\gamma_i$ defines the peak width, $\Omega_i$ defines the center frequency, and $\beta$ is the inverse temperature. The PC645 system Hamiltonian was constructed by Mirkovich et. al \cite{Mirkovic:2007bt} assuming an identical spectral density for all sites with a reorganization energy of 478.24 $cm^{-1}$. In order to combine the Mirkovich Hamiltonian with our spectral densities, we need to modify the diagonal elements, which denote the excited state energy of each chromophore plus the reorganization energy of that chromophore as defined above. To account for the reorganization energies of our unique spectral densities, we subtract off 478.24 $cm^{-1}$ from each site energy and add the corresponding reorganization energy of that bilin's spectral density, preserving the underlying $\epsilon_m^0$ values obtained previously.

\section*{Spectral Densities}
\addcontentsline{toc}{section}{Spectral Densities}
\hspace{\parindent} While HEOM is numerically exact, large reorganization energy values require substantial hierarchy depth to ensure convergence. The particularly large reorganization energies of the MBVs require a depth of Nmax=6 for all full-system calculations presented in this manuscript, which restricts us to including a total of 24 bath modes over all eight sites and necessitates coarse graining of our spectral densities. We construct four classes of abridged spectral densities for each bilin that include successively fewer peaks. Construction is guided by four metrics: 

\begin{itemize}
\item L1 and L2 norms of error between the abridged and unabridged spectral densities
\item the relative distribution of reorganization energy along the frequency axis, e.g. how much of the reorganization energy is contained between 0 and 200 $cm^{-1}$ versus between 200 and 400 $cm^{-1}$, etc
\item monomer absorption and fluorescence lineshapes
\item three-site population dynamics, including the core DBV bilins along with each other site in turn
\end{itemize}
We note that no experimental data of any kind was referenced during spectral density construction. Leveraging all five of these metrics simultaneously allows us to systematically and optimally reduce complexity while preserving essential observables given the constraints.

We define the first and most accurate set of spectral densities as Class~1, and these include either eight or nine peaks depending on the bilin. Class~1 spectral densities have enough peaks to accurately capture the majority of the sharp features observed in the unabridged spectral densities, but contain too many bath modes to be applied to full-system calculations given the numerical complexity of the hierarchical equations of motion exciton dynamics method. Class 2 spectral densities include either five or six peaks depending on the bilin and therefore can only accurately represent some of the sharp features observed in the unabridged spectral densities. Despite the limitations, simulations with Class 2 spectral densities are able to accurately reproduce absorption, fluorescence, and population dynamics obtained from simulations with the more accurate Class~1 spectral densities. Class~3 and Class~4 spectral densities contain three and two peaks, respectively, and cannot accurately capture any of the sharp features observed in the unabridged spectral densities. Spectral densities of all four classes for $PCB82_C$ are shown in Figure S1, while spectral density parameters for all classes and all bilins are given below in Table S1, S2, S3, and S4.

\begin{figure*}[t!]
\begin{center}
\includegraphics[width=0.95\textwidth]{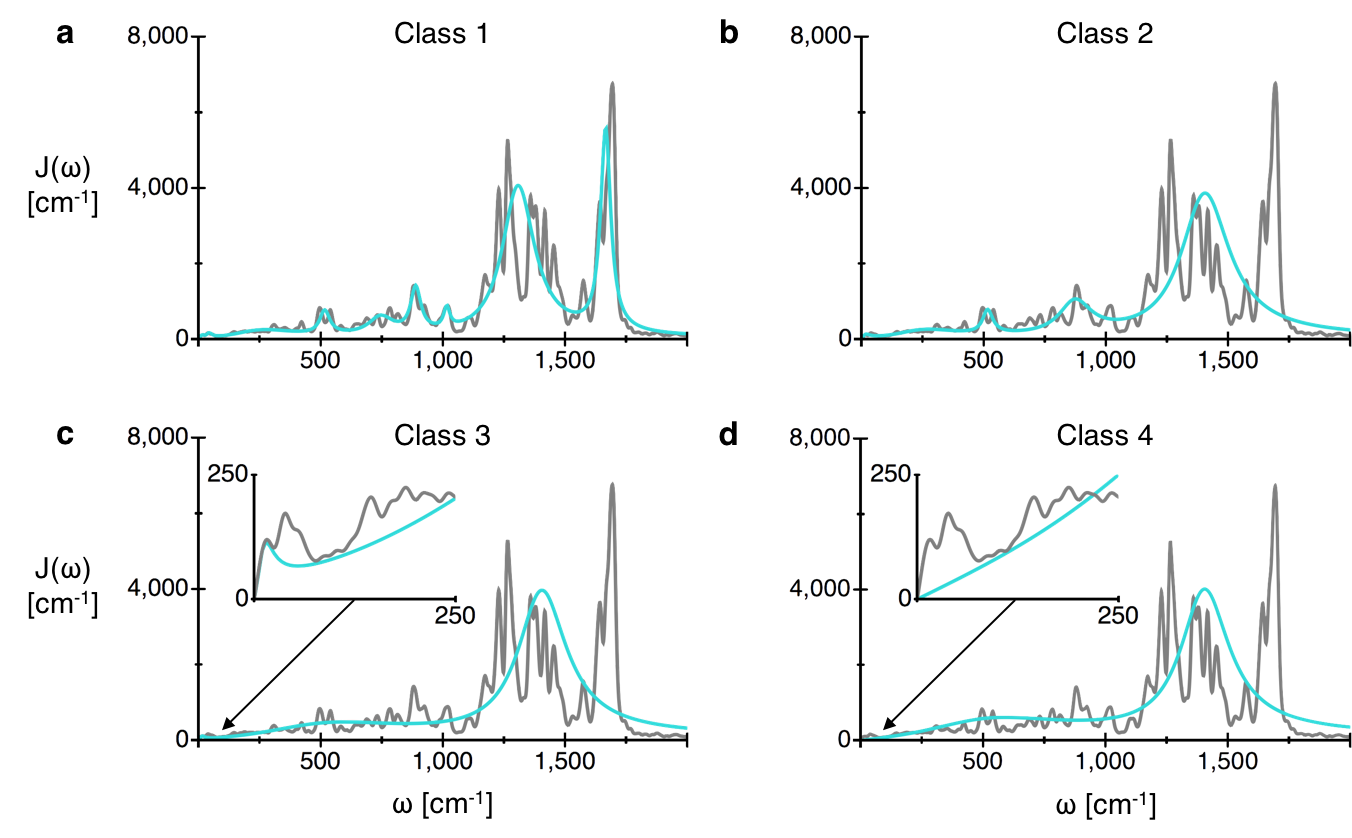}
\caption{\textbf{Four spectral density classes for PCB82C. a}, Class~1, nine peaks. \textbf{b}, Class 2, five peaks. \textbf{c}, Class 3, three peaks. \textbf{d}, Class 4, two peaks.}
\end{center}
\end{figure*}

\begin{figure*}[t!]
\begin{center}
\includegraphics[width=0.95\textwidth]{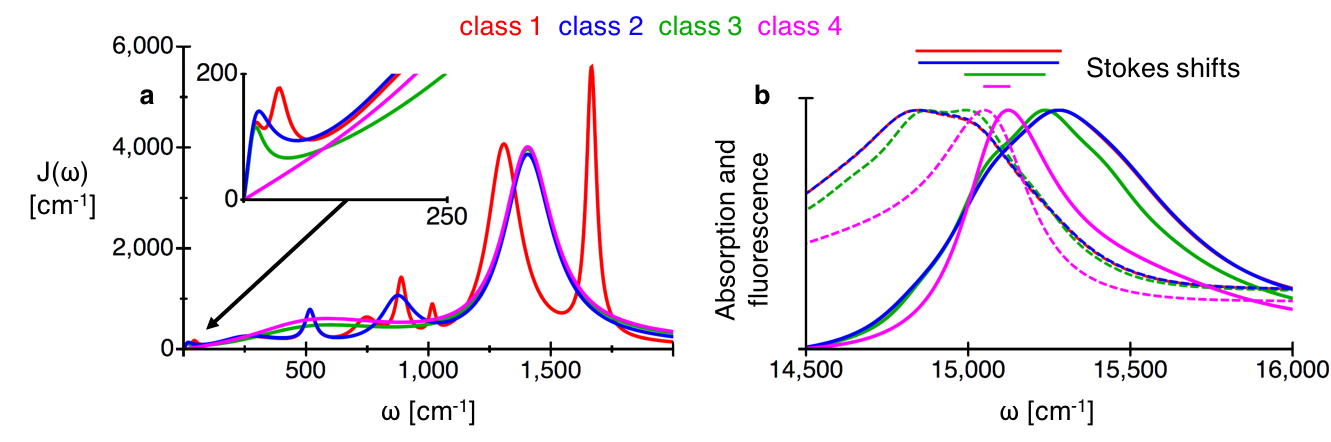}
\caption{\textbf{Monomer absorption and fluorescence. a}, All four classes of spectral densities for $PCB82_C$ with Class~1 (nine peaks) shown in red, Class 2 (5 peaks) shown in blue, Class 3 (3 peaks) shown in green, and Class 4 (two peaks) shown in magenta. \textbf{b}, Monomer absorption (solid) and fluorescence (dashed) calculated with HEOM using the four classes of spectral densities. Stokes shifts are depicted graphically above.}
\end{center}
\end{figure*}

We construct Class~1 spectral densities by trying to minimize L1 and L2 error with respect to the unabridged spectral densities while also preserving the relative distribution of reorganization energy. Class~1 spectral densities are too complex for use in full-system calculations, but we can use them to run monomer HEOM calculations up to a hierarchy depth of Nmax=12 for precise convergence. As seen in Figure S2, for $PCB82_C$, monomer simulations with the Class 2 spectral density (blue) are able to reproduce absorption and fluorescence lineshapes and Stokes shift obtained from simulations with the Class~1 spectral density (red). While the Class 3 and Class 4 spectral densities cannot reproduce the lineshape or Stokes shift, the addition of a peak near zero frequency in Class 3 compared with Class 4 dramatically improves both observables, as shown in Figure S2. This demonstrates the importance of accurately capturing the pure dephasing rate, which is defined as the spectral density slope at zero frequency, since it directly impacts the optical dephasing time which influences Stokes shift. We were able to obtain Class 2 spectral densities that accurately reproduced both lineshape and Stokes shift obtained from Class~1 spectral densities for all eight bilins. As seen in Figure S3, while we found population dynamics to be less sensitive to spectral density structure compared with spectroscopy, MBV bilin population dynamics in particular remained susceptible to changes in the spectral density structure. Therefore, full system population dynamics and population flux simulations presented in the main text use Class 2 spectral densities for MBVs and Class 4 spectral densities for all other bilins.

\begin{figure*}[t!]
\begin{center}
\includegraphics[width=0.9\textwidth]{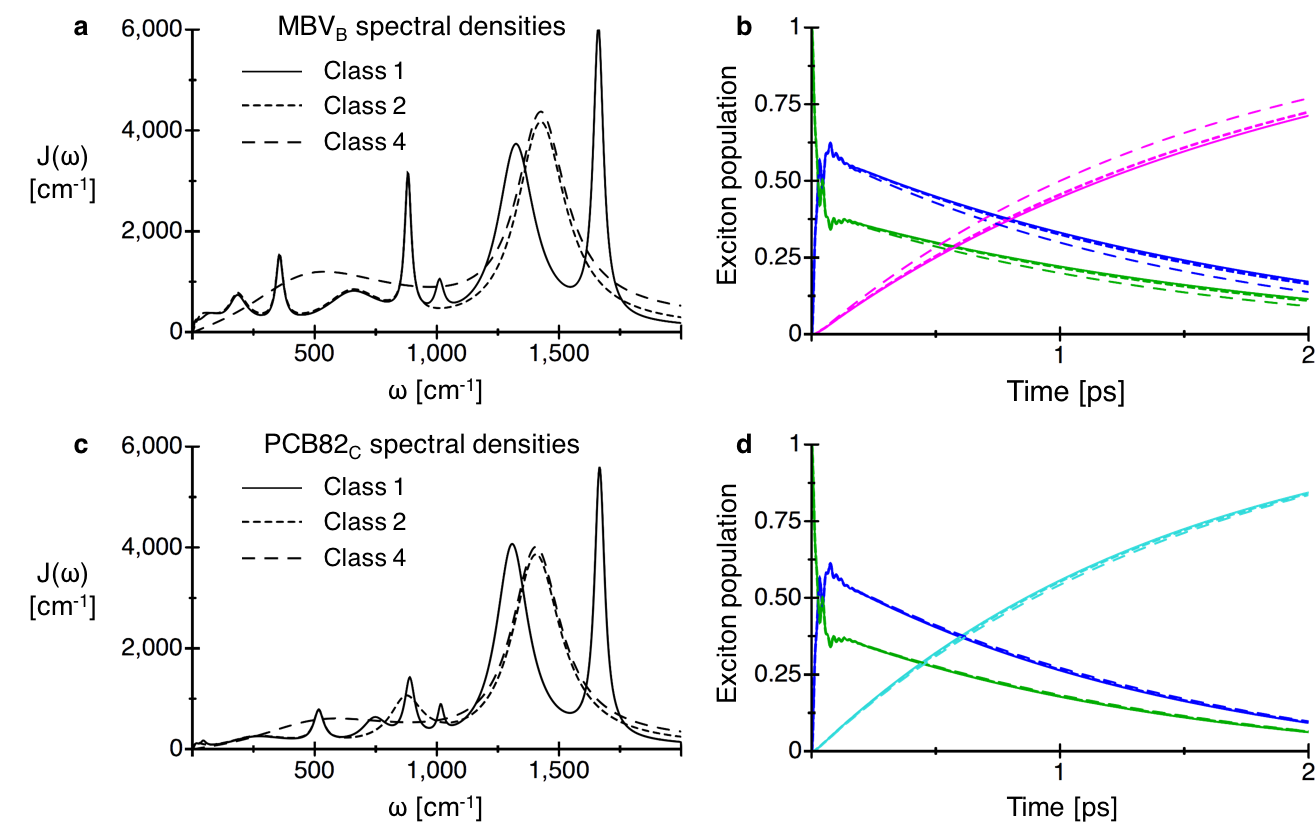}
\caption{\textbf{Three-site population dynamics.} We note that Class~3 have dynamics identical to Class~4 spectral densities. \textbf{a}, Three classes of $MBV_B$ spectral densities with Class~1 shown in solid, Class 2 shown in short dashed, and Class 4 shown in long dashed. \textbf{b}, Three-site population dynamics from the DBV core (blue and green) to $MBV_B$ (magenta) using the three different classes of spectral density for the $MBV_B$ while Class 4 spectral densities were used for the DBV core. \textbf{c}, Three classes of $PCB82_C$ spectral densities with Class~1 shown in solid, Class 2 shown in short dashed, and Class 4 shown in long dashed. \textbf{d}, Three-site population dynamics from the DBV core (blue and green) to $PCB82_C$ (cyan) using the three different classes of spectral density for the $PCB82_C$ while Class 4 spectral densities were used for the DBV core.
}
\end{center}
\end{figure*}

\begin{table}
\centering
% \begin{tabular}{ c | c | c || c | c | c }
\begin{tabular}{ c  c  c | c  c  c }
\multicolumn{3}{c}{} & \multicolumn{3}{c}{} \\ 
         & $\mathbf{DBV_C}$   &           &          & $\mathbf{DBV_D}$   &             \\
$\lambda (cm^{-1})$ & $\gamma (cm^{-1})$ & $\Omega (cm^{-1})$  & $\lambda (cm^{-1})$ & $\gamma (cm^{-1})$ & $\Omega (cm^{-1})$    \\ \hline 
 95.8 &  7.50 &   5.95 & 114.0 & 10.50 &    4.0 \\
148.2 & 90.00 &  110.0 & 120.0 & 70.00 &   80.0 \\
 63.0 & 16.00 &  352.0 &  52.0 & 50.00 &  310.0 \\
 50.0 & 37.00 &  540.0 &  24.2 & 13.00 &  480.0 \\
 40.0 & 40.00 &  740.0 &  15.4 & 12.00 &  599.0 \\
 71.6 & 55.00 &  945.0 &  70.0 & 55.00 &  898.0 \\
183.6 & 80.00 & 1330.0 & 186.6 & 80.00 & 1370.0 \\
 96.0 & 30.00 & 1620.0 &  90.4 & 25.00 & 1660.0 \\
\noalign{\vskip 2.5mm}  
         & $\mathbf{MBV_A}$   &           &          & $\mathbf{MBV_B}$   &             \\
$\lambda (cm^{-1})$ & $\gamma (cm^{-1})$ & $\Omega (cm^{-1})$  & $\lambda (cm^{-1})$ & $\gamma (cm^{-1})$ & $\Omega (cm^{-1})$    \\ \hline
107.4 &   6.50 &    5.0 & 162.0 &  10.20 &    7.0 \\
268.0 &  40.00 &   60.0 & 200.0 &  50.00 &   53.0 \\
264.0 & 100.00 &  255.0 & 131.0 &  43.00 &  180.0 \\
 48.0 &  30.00 &  530.0 &  66.0 &  18.00 &  355.0 \\
200.0 &  80.00 &  800.0 & 150.6 & 140.00 &  643.0 \\
 20.0 &  14.00 & 1010.0 &  50.0 &  16.00 &  881.0 \\
376.0 &  90.00 & 1300.0 &  12.0 &  20.00 & 1010.0 \\
160.0 &  25.00 & 1640.0 & 245.8 &  90.00 & 1320.0 \\
      &        &        &  86.0 &  25.00 & 1660.0 \\
\noalign{\vskip 2.5mm}  
         & $\mathbf{PCB158_C}$ &           &          & $\mathbf{PCB158_D}$ &             \\
$\lambda (cm^{-1})$ & $\gamma (cm^{-1})$ & $\Omega (cm^{-1})$  & $\lambda (cm^{-1})$ & $\gamma (cm^{-1})$ & $\Omega (cm^{-1})$    \\ \hline
128.0 &  15.00 &   14.0 & 154.0 &   7.50 &    6.0 \\
 80.0 &  30.00 &  105.0 &  60.8 &  34.00 &   69.0 \\
 46.0 &  18.00 &  197.0 & 110.6 &  72.00 &  275.0 \\
118.0 &  55.00 &  330.0 &  24.0 &  20.00 &  545.0 \\
 56.0 &  25.00 &  550.0 & 150.0 &  84.00 &  725.0 \\
185.2 &  88.00 &  780.0 &  24.0 &  15.00 &  888.0 \\
 14.0 &   9.00 & 1014.0 &  12.0 &  10.00 & 1005.0 \\
242.2 &  80.00 & 1300.0 & 332.6 &  80.00 & 1260.0 \\
104.6 &  23.00 & 1635.0 & 158.0 &  60.00 & 1600.0 \\
\noalign{\vskip 2.5mm}  
         & $\mathbf{PCB82_C}$ &           &          & $\mathbf{PCB82_D}$ &             \\
$\lambda (cm^{-1})$ & $\gamma (cm^{-1})$ & $\Omega (cm^{-1})$  & $\lambda (cm^{-1})$ & $\gamma (cm^{-1})$ & $\Omega (cm^{-1})$    \\ \hline
 79.6 &  10.50 &    9.5 &  63.4 &   8.00 &    5.0 \\
 30.4 &  12.00 &   42.0 &  92.0 &  80.00 &   70.0 \\
131.2 & 150.00 &  223.0 &  28.0 &  24.50 &  220.0 \\
 30.0 &  25.00 &  515.0 &  70.0 &  80.00 &  488.0 \\
 40.0 &  62.00 &  743.0 &  58.0 &  65.00 &  810.0 \\
 33.0 &  25.00 &  888.0 &  16.0 &  16.00 & 1009.0 \\
  8.0 &  15.00 & 1015.0 & 220.0 &  80.00 & 1305.0 \\
244.8 &  80.00 & 1305.0 &  80.0 &  22.00 & 1675.0 \\
 80.0 &  25.00 & 1665.0 &       &        &        \\
\end{tabular}
\caption{Class~1 spectral density parameters.}
\end{table}

\begin{table}
\centering
% \begin{tabular}{ c | c | c || c | c | c }
\begin{tabular}{ c  c  c | c  c  c }
\multicolumn{3}{c}{} & \multicolumn{3}{c}{} \\ 
         & $\mathbf{DBV_C}$   &           &          & $\mathbf{DBV_D}$   &             \\
$\lambda (cm^{-1})$ & $\gamma (cm^{-1})$ & $\Omega (cm^{-1})$  & $\lambda (cm^{-1})$ & $\gamma (cm^{-1})$ & $\Omega (cm^{-1})$    \\ \hline 
244.0 &  90.00 &    0.0 & 234.0 &  80.00 &    0.0 \\
 63.0 &  16.00 &  352.0 &  52.0 &  40.00 &  310.0 \\
 50.0 &  37.00 &  540.0 &  20.2 &  13.00 &  480.0 \\
 40.0 &  40.00 &  740.0 &  15.4 &  12.00 &  599.0 \\
 71.6 &  55.00 &  945.0 &  70.0 &  55.00 &  898.0 \\
279.6 & 120.00 & 1400.0 & 281.0 & 120.00 & 1420.0 \\
\noalign{\vskip 2.5mm}  
         & $\mathbf{MBV_A}$   &           &          & $\mathbf{MBV_B}$   &             \\
$\lambda (cm^{-1})$ & $\gamma (cm^{-1})$ & $\Omega (cm^{-1})$  & $\lambda (cm^{-1})$ & $\gamma (cm^{-1})$ & $\Omega (cm^{-1})$    \\ \hline 
375.4 &  50.00 &    0.0 & 362.0 &  50.00 &    0.0 \\
264.0 & 100.00 &  255.0 & 131.0 &  43.00 &  180.0 \\
 48.0 &  30.00 &  530.0 &  66.0 &  18.00 &  355.0 \\
 40.0 &  18.00 &  695.0 & 150.6 & 140.00 &  643.0 \\
180.0 &  80.00 &  900.0 &  50.0 &  16.00 &  881.0 \\
536.0 & 120.00 & 1400.0 & 343.8 & 120.00 & 1420.0 \\
\noalign{\vskip 2.5mm}  
         & $\mathbf{PCB158_C}$ &           &          & $\mathbf{PCB158_D}$ &             \\
$\lambda (cm^{-1})$ & $\gamma (cm^{-1})$ & $\Omega (cm^{-1})$  & $\lambda (cm^{-1})$ & $\gamma (cm^{-1})$ & $\Omega (cm^{-1})$    \\ \hline 
128.0 &  15.00 &   14.0 & 214.8 &  34.00 &    0.0 \\
 80.0 &  30.00 &  105.0 & 110.6 &  72.00 &  275.0 \\
164.0 &  55.00 &  300.0 &  24.0 &  20.00 &  545.0 \\
 56.0 &  25.00 &  550.0 & 174.0 &  84.00 &  750.0 \\
199.2 & 100.00 &  810.0 &  12.0 &  10.00 & 1005.0 \\
346.8 & 120.00 & 1420.0 & 490.6 & 140.00 & 1420.0 \\
\noalign{\vskip 2.5mm}  
         & $\mathbf{PCB82_C}$ &           &          & $\mathbf{PCB82_D}$ &             \\
$\lambda (cm^{-1})$ & $\gamma (cm^{-1})$ & $\Omega (cm^{-1})$  & $\lambda (cm^{-1})$ & $\gamma (cm^{-1})$ & $\Omega (cm^{-1})$    \\ \hline 
110.0 &  15.00 &    9.5 &  63.4 &   8.00 &    5.0 \\
131.2 & 150.00 &  223.0 &  92.0 &  80.00 &   70.0 \\
 30.0 &  25.00 &  515.0 &  28.0 &  24.50 &  220.0 \\
 81.0 &  80.00 &  870.0 & 144.0 & 180.00 &  700.0 \\
324.8 & 120.00 & 1400.0 & 300.0 & 120.00 & 1400.0 \\
\end{tabular}
\caption{Class 2 spectral density parameters.}
\end{table}

\begin{table}
\centering
% \begin{tabular}{ c | c | c || c | c | c }
\begin{tabular}{ c  c  c | c  c  c }
\multicolumn{3}{c}{} & \multicolumn{3}{c}{} \\ 
         & $\mathbf{DBV_C}$   &           &          & $\mathbf{DBV_D}$   &             \\
$\lambda (cm^{-1})$ & $\gamma (cm^{-1})$ & $\Omega (cm^{-1})$  & $\lambda (cm^{-1})$ & $\gamma (cm^{-1})$ & $\Omega (cm^{-1})$    \\ \hline 
 95.8 &   7.50 &   5.95 & 114.0 &  10.50 &    4.0 \\
372.8 & 250.00 &  400.0 & 267.0 & 350.00 &  400.0 \\
279.6 & 120.00 & 1400.0 & 291.6 & 120.00 & 1420.0 \\
\noalign{\vskip 2.5mm}  
         & $\mathbf{MBV_A}$   &           &          & $\mathbf{MBV_B}$   &             \\
$\lambda (cm^{-1})$ & $\gamma (cm^{-1})$ & $\Omega (cm^{-1})$  & $\lambda (cm^{-1})$ & $\gamma (cm^{-1})$ & $\Omega (cm^{-1})$    \\ \hline 
107.4 &   6.50 &    5.0 & 162.0 &  10.20 &    7.0 \\
800.0 & 350.00 &  400.0 & 609.6 & 350.00 &  400.0 \\
536.0 & 120.00 & 1400.0 & 331.8 & 120.00 & 1420.0 \\
\noalign{\vskip 2.5mm}  
         & $\mathbf{PCB158_C}$ &           &          & $\mathbf{PCB158_D}$ &             \\
$\lambda (cm^{-1})$ & $\gamma (cm^{-1})$ & $\Omega (cm^{-1})$  & $\lambda (cm^{-1})$ & $\gamma (cm^{-1})$ & $\Omega (cm^{-1})$    \\ \hline 
128.0 &  15.00 &   14.0 & 154.0 &   7.50 &    6.0 \\
499.2 & 350.00 &  400.0 & 381.4 & 350.00 &  450.0 \\
346.8 & 120.00 & 1420.0 & 490.6 & 140.00 & 1420.0 \\
\noalign{\vskip 2.5mm}  
         & $\mathbf{PCB82_C}$ &           &          & $\mathbf{PCB82_D}$ &             \\
$\lambda (cm^{-1})$ & $\gamma (cm^{-1})$ & $\Omega (cm^{-1})$  & $\lambda (cm^{-1})$ & $\gamma (cm^{-1})$ & $\Omega (cm^{-1})$    \\ \hline 
 79.6 &  10.50 &    9.5 &  63.4 &   8.00 &    5.0 \\
272.6 & 350.00 &  450.0 & 264.0 & 350.00 &  450.0 \\
324.8 & 120.00 & 1400.0 & 300.0 & 120.00 & 1400.0 \\
\end{tabular}
\caption{Class 3 spectral density parameters.}
\end{table}

\begin{table}
\centering
\begin{tabular}{ c  c  c | c  c  c }
\multicolumn{3}{c}{} & \multicolumn{3}{c}{} \\ 
         & $\mathbf{DBV_C}$   &           &          & $\mathbf{DBV_D}$   &             \\
$\lambda (cm^{-1})$ & $\gamma (cm^{-1})$ & $\Omega (cm^{-1})$  & $\lambda (cm^{-1})$ & $\gamma (cm^{-1})$ & $\Omega (cm^{-1})$    \\ \hline 
468.6 & 250.00 &  400.0 & 381.0 & 350.00 &  400.0 \\
279.6 & 120.00 & 1400.0 & 291.6 & 120.00 & 1420.0 \\
\noalign{\vskip 2.5mm}  
         & $\mathbf{MBV_A}$   &           &          & $\mathbf{MBV_B}$   &             \\
$\lambda (cm^{-1})$ & $\gamma (cm^{-1})$ & $\Omega (cm^{-1})$  & $\lambda (cm^{-1})$ & $\gamma (cm^{-1})$ & $\Omega (cm^{-1})$    \\ \hline 
907.4 & 350.00 &  400.0 & 771.6 & 350.00 &  400.0 \\
536.0 & 120.00 & 1400.0 & 331.8 & 120.00 & 1420.0 \\
\noalign{\vskip 2.5mm}  
         & $\mathbf{PCB158_C}$ &           &          & $\mathbf{PCB158_D}$ &             \\
$\lambda (cm^{-1})$ & $\gamma (cm^{-1})$ & $\Omega (cm^{-1})$  & $\lambda (cm^{-1})$ & $\gamma (cm^{-1})$ & $\Omega (cm^{-1})$    \\ \hline 
627.2 & 350.00 &  400.0 & 535.4 & 350.00 &  450.0 \\
346.8 & 120.00 & 1420.0 & 490.6 & 140.00 & 1420.0 \\
\noalign{\vskip 2.5mm}  
         & $\mathbf{PCB82_C}$ &           &          & $\mathbf{PCB82_D}$ &             \\
$\lambda (cm^{-1})$ & $\gamma (cm^{-1})$ & $\Omega (cm^{-1})$  & $\lambda (cm^{-1})$ & $\gamma (cm^{-1})$ & $\Omega (cm^{-1})$    \\ \hline 
352.2 & 350.00 &  450.0 & 327.4 & 350.00 &  450.0 \\
324.8 & 120.00 & 1400.0 & 300.0 & 120.00 & 1400.0 \\
\end{tabular}
\caption{Class 4 spectral density parameters.}
\end{table}

\section*{Spectroscopy}
\addcontentsline{toc}{section}{Spectroscopy}
\hspace{\parindent} {\it{Ab initio}} transition dipole moments for each bilin, necessary for accurate spectroscopic simulations, were obtained from TDDFT calculations and averaged over bilin trajectories. Our eight transition dipole vectors are given in Table S5.

\begin{table}
\begin{center}
\begin{tabular}{ c | c | c | c}
\multicolumn{3}{c}{} \\ 
\textbf{Bilin} & $\mathbf{L_x}$ & $\mathbf{L_y}$ & $\mathbf{L_z}$  \\ \hline 
$DBV_C$    & 1.316 &  3.062 &  3.082 \\
$DBV_D$    & 0.794 &  3.260 &  2.604 \\
$MBV_A$    & 0.882 &  3.535 & -2.359 \\
$MBV_B$    & 0.072 & -3.517 &  2.947 \\
$PCB158_C$ & 3.414 &  0.195 &  2.944 \\
$PCB158_D$ & 3.304 & -1.780 & -2.438 \\
$PCB82_C$  & 0.935 & -1.908 &  4.243 \\
$PCB82_D$  & 1.949 &  0.365 & -4.359 \\
\end{tabular}
\caption{Transition dipole vectors given in atomic units.}
\end{center}
\end{table}

Inhomogeneous broadening was accounted for in absorption and fluorescence simulations by averaging 100 calculations in which each bilin excitation energy was perturbed by a value drawn randomly from a Gaussian distribution with a width of 150 $cm^{-1}$. This width was selected to be approximately the average of the inhomogeneous line broadening values used by Mirkovich et. al \cite{Mirkovic:2007bt}.

Fluorescence results presented in the main text are Boltzmann weighted sums of monomer fluorescence calculations run with HEOM at a hierarchy depth of Nmax=12. The additional computational complexity of fluorescence compared to absorption prevents full system simulations from achieving convergence in the hierarchy. Calculations including only the four lowest energy bilins, the PCB158s and PCB82s, are feasible and reasonably accurate given that fluorescence spectra are dominated by the response of low energy sites. We observed that four-site calculations are indistinguishable from a Boltzmann weighted sum of four monomer lineshapes. Therefore, we present monomer-based results that allow larger Nmax values.

Absorption and fluorescence spectra presented in the main text can be seen to go slightly below zero amplitude, which is unphysical. We expect negative features arise due to the high-temperature approximation made by the QMaster implementation of HEOM. In order to ensure that negative features are not impacting the overall lineshape, we compared HEOM monomer absorption and fluorescence spectra to Kubo lineshapes, which are exact for monomers, shown in Figure S4. Monomer calculations employed Class 1 spectral densities, and the HEOM calculations were run at a hierarchy depth of Nmax=12. While HEOM lineshapes appear to be slightly too narrow, their features remain minimally perturbed by negative features.

\begin{figure*}[t!]
\begin{center}
\includegraphics[width=0.9\textwidth]{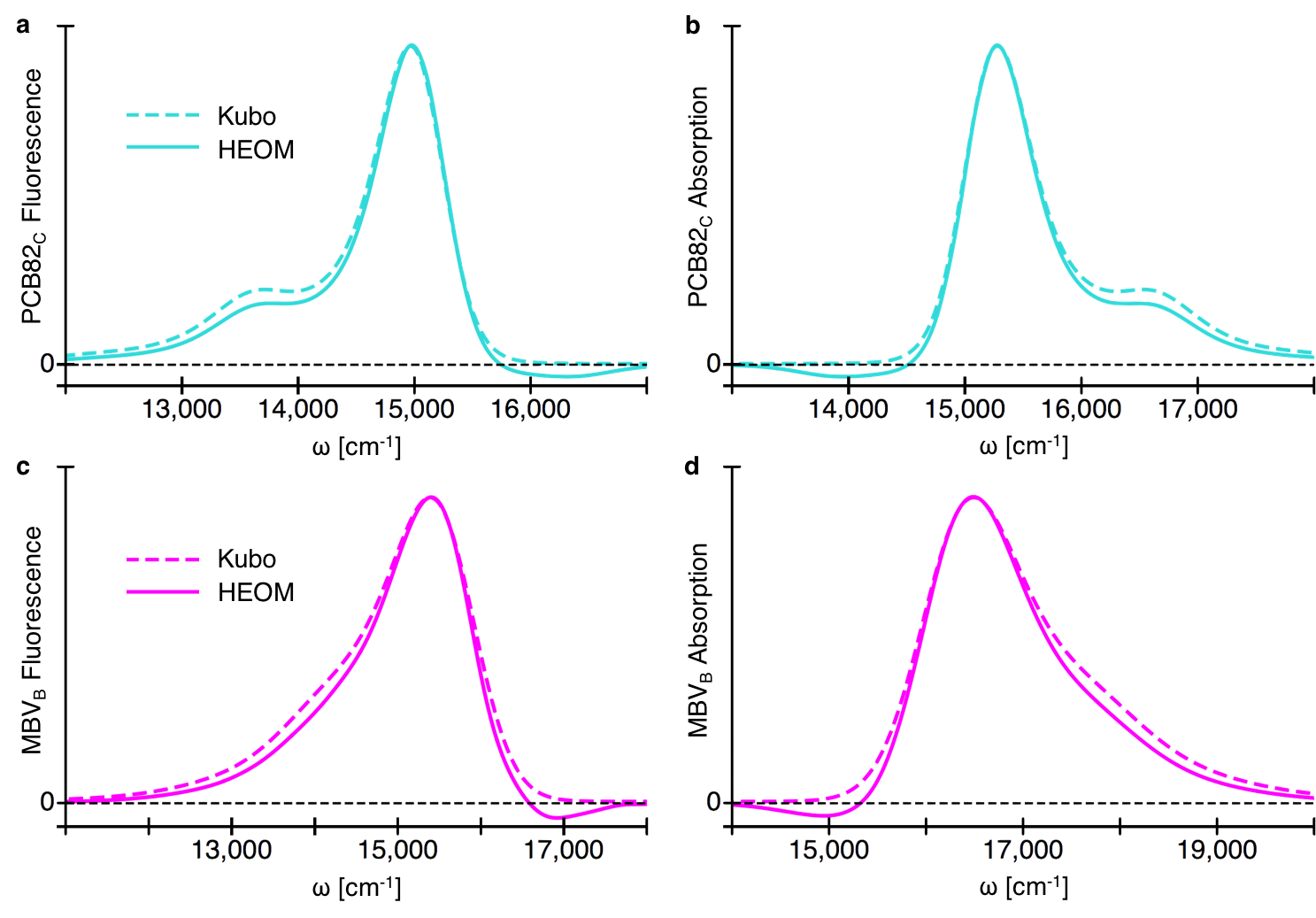}
\caption{\textbf{Comparing monomer lineshapes.} Monomer Kubo absorption and fluorescence (dashed) are compared to monomer HEOM absorption and fluorescence (solid) for two representative bilins, $MBV_B$ (magenta) and $PCB82_C$ (cyan).}
\end{center}
\end{figure*}

Absorption and fluorescence spectra are highly sensitive to spectral density structures and pigment reorganization energies. Due to experimental constraints, spectral densities of the lowest energy bilins, such as would be extracted from fluorescence line narrowing, are often used for all pigments. However, if we employ $PCB82_C$ spectral densities for all pigments, we obtain an absorption spectrum with a full-width at half maximum 25\% too small, as seen in Figure S5. The decreased spectral width is due to the reduced reorganization energy of the high-energy MBVs and DBVs, while the Stokes shift remains accurate given that the reorganization energies of the PCB82s remain essentially unchanged. In order to present a fair comparison with our non-identical spectra, Class 4 $PCB82_C$ spectral densities were used for MBVs and DBVs, Class 3 $PCB82_C$ spectral densities were used for PCB158s, and Class 2 $PCB82_C$ spectral densities were used for PCB82s.

\begin{figure*}[t!]
\begin{center}
\includegraphics[width=0.75\textwidth]{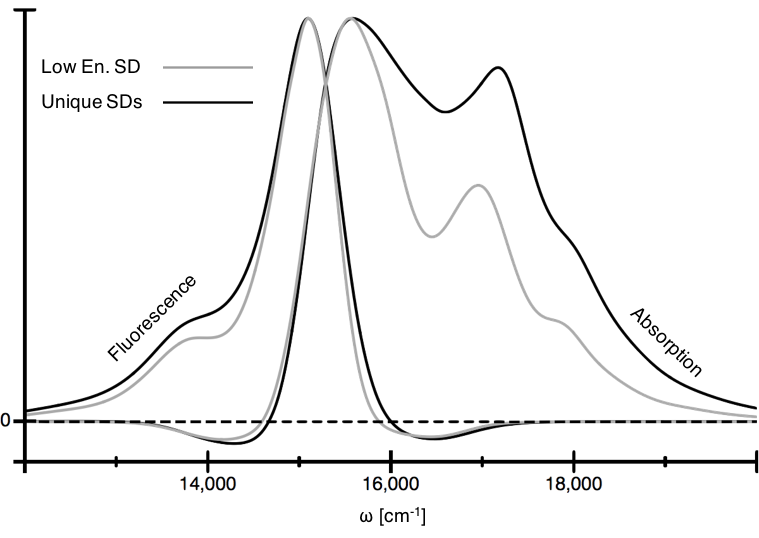}
\caption{\textbf{The sensitivity of linear spectra to spectral densities: low energy SDs} Absorption and fluorescence spectra using spectral densities from the lowest energy bilin $PCB82_C$ (gray) compared with those using unique spectral densities (black).}
\end{center}
\end{figure*}

In contrast, if we employ the average of the eight bilin spectral densities for all pigments, we obtain an absorption spectrum that overestimates the Stokes shift by 55\%, as seen in Figure S6. The increased Stokes shift follows from the increase in the reorganization energy of our lowest energy pigments when using the average spectral density. We note that average spectral densities were made by averaging the eight raw bilin spectral densities and then using the same construction procedure to obtain the four spectral density classes as detailed previously. Once again, in order to present a fair comparison with our non-identical spectra, Class 4 average spectral densities were used for MBVs and DBVs, Class 3 average spectral densities were used for PCB158s, and Class 2 average spectral densities were used for PCB82s.

\begin{figure*}[t!]
\begin{center}
\includegraphics[width=0.9\textwidth]{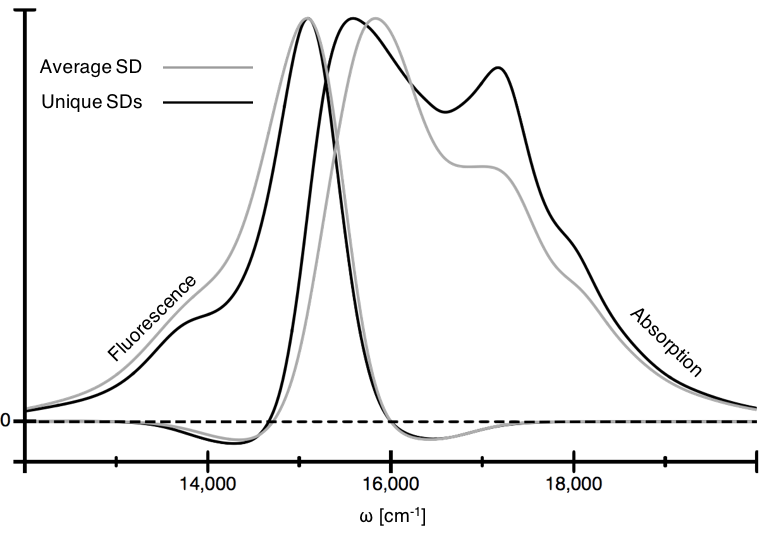}
\caption{\textbf{The sensitivity of linear spectra to spectral densities: average SDs} Absorption and fluorescence spectra using average spectral densities (gray) compared with those using unique spectral densities (black).}
\end{center}
\end{figure*}

We note that in both of the examples presented, Hamiltonian site energies were adjusted according to the reorganization energies in order to continue to preserve the underlying $\epsilon_m^0$ values. While many different sets of spectral densities could in principle yield reasonable absorption and fluorescence spectra, the significant spectroscopic error introduced when employing two common approximations provides further evidence that our spectral densities and their associated reorganization energies have accurately captured physically relevant details of individual bilin vibrational environments in PC645. Finally, we note that for both the examples presented here and the spectra presented in the main text, an additional 85 $cm^{-1}$ was added to all site energies to ensure alignment with the experimental fluorescence peak. This has no impact on the Stokes shift or the absorption lineshape that we depend on to validate our spectral densities.

\section*{Dynamics}
\addcontentsline{toc}{section}{Dynamics}
\hspace{\parindent} Figure S7 shows one picosecond of full system population dynamics calculated at a hierarchy depth of Nmax=6. The exciton is initially localized on the highest energy bilin, $DBV_D$ (green), and can be seen to rapidly delocalize over the DBV core (blue and green) before down-converting to the low energy PCB82s (cyan and red). The minimal populations of the intermediate MBVs (magenta and brown) represent an improvement over previous simulation of PC645 exciton dynamics \cite{Huo2011a} given the better agreement with experimental transient absorption measurements \cite{Marin2011a}.

\begin{figure*}[t!]
\begin{center}
\includegraphics[width=0.9\textwidth]{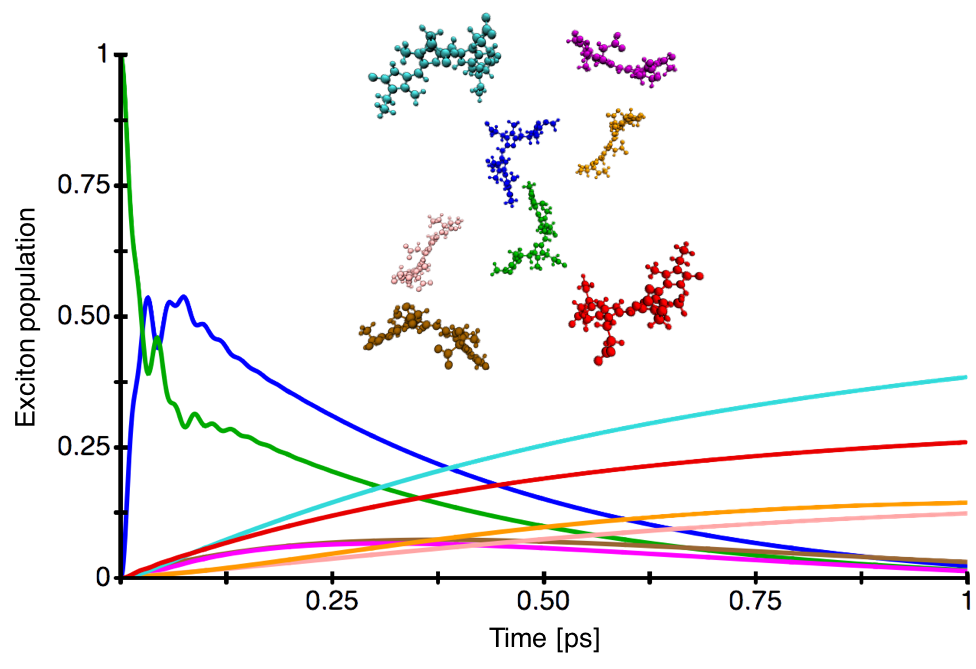}
\caption{\textbf{Full system population dynamics.}  One picosecond of population dynamics calculated with Class 2 spectral densities (six peaks) on the MBVs (brown and magenta) and Class 4 spectral densities (two peaks) on all other sites.}
\end{center}
\end{figure*}

We perform four-site HEOM population dynamics simulations including the DBV core and the low energy PCB82s in order to isolate the DBV to PCB transfer rate. The populations of the two DBVs and the two PCB82s are summed separately in order to obtain a dimer-like system from which we extract rates as per Dijkstra et. al \cite{Dijkstra:2015er}. Summed populations obtained from HEOM calculations are compared to populations obtained from resulting rate equations in Figure S8 and are found to agree with minimal error.

\begin{figure*}[t!]
\begin{center}
\includegraphics[width=0.9\textwidth]{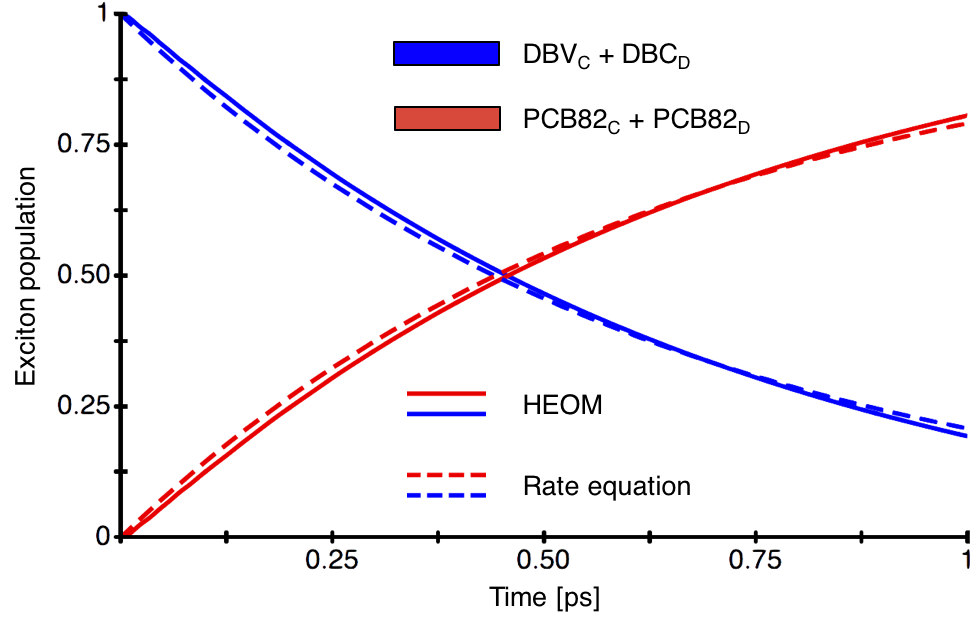}
\caption{\textbf{DBV to PCB82 populations: HEOM vs rate equation.} Summed DBV populations (blue) and summed PCB82 populations (red) obtained from HEOM calculations (solid) are compared with the summed populations given by a rate equation (dashed) that employs the extracted rate of 1.57 $ps^{-1}$.}
\end{center}
\end{figure*}

We calculate rates in this way at a range of scaled reorganization energies and using two different classes of spectral densities to further demonstrate the importance of our large reorganization values and of resolved spectral density features for accurate rate prediction. As seen in Figure S9, the rate in best agreement with experiment is obtained at our full reorganization energy and with the more accurate Class 2 spectral densities.

While rate calculations presented here were run at a hierarchy depth of Nmax=6, our four-site population dynamics no longer include the bilins with the largest reorganization energies, and thus we determined that they converged at a hierarchy depth of Nmax=5. The significantly reduced computational cost at this depth allowed us to use Class 1 spectral densities to simulate down-conversion rates presented in the main text while maintaining strict convergence. We note that the rate of down-conversion obtained from simulations with Class 1 spectral densities differs by less than 0.015 $ps^{-1}$ compared with the rate obtained from simulations with Class 2 spectral densities.

\begin{figure*}[t!]
\begin{center}
\includegraphics[width=0.9\textwidth]{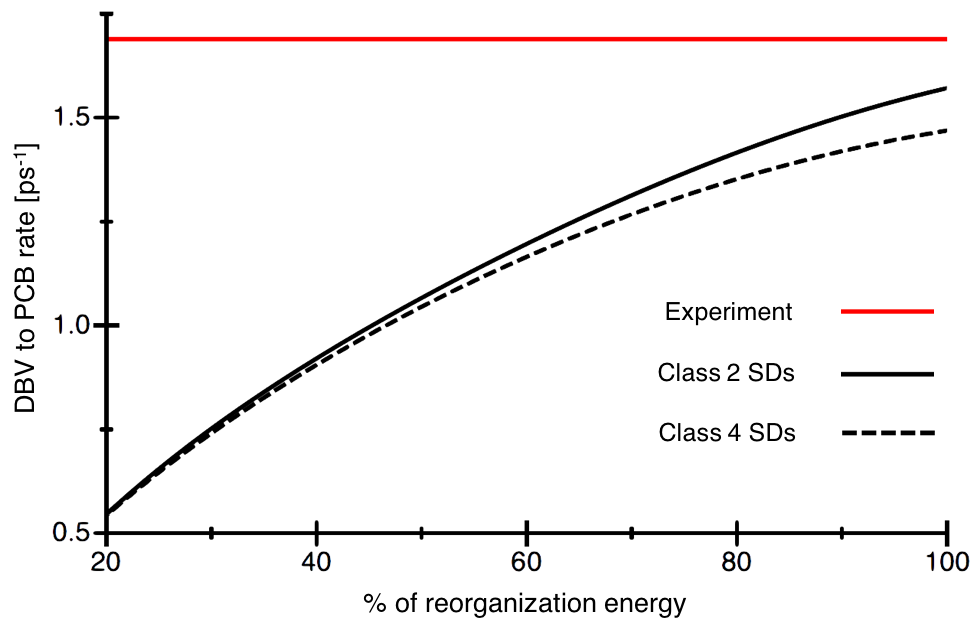}
\caption{\textbf{DBV to PCB transfer rate as a function of reorganization energy.} EET rate as a function of reorganization energy scaling extracted from four-site HEOM calculations. The experimental rate is shown in red, HEOM simulations with Class 2 spectral densities (five peaks) shown in solid black, and HEOM simulations with Class 4 spectral densities (two peaks) shown in dashed black.}
\end{center}
\end{figure*}

F{\"o}rster overlaps presented in the main text rely on the accuracy of Kubo fluorescence for the tightly bound DBV core. However, Kubo lineshapes fail to describe the time-dependent dynamic localization process captured by HEOM, as shown in Figure S10. The additional Stokes shift increases overlap with the PCB82 absorption by 12\%. The increased rate is mitigated by an estimated 10\% decrease of $\vert V_{eff}^2 \vert$ as a result of dynamic localization in the DBV core.

\begin{figure*}[t!]
\begin{center}
\includegraphics[width=0.9\textwidth]{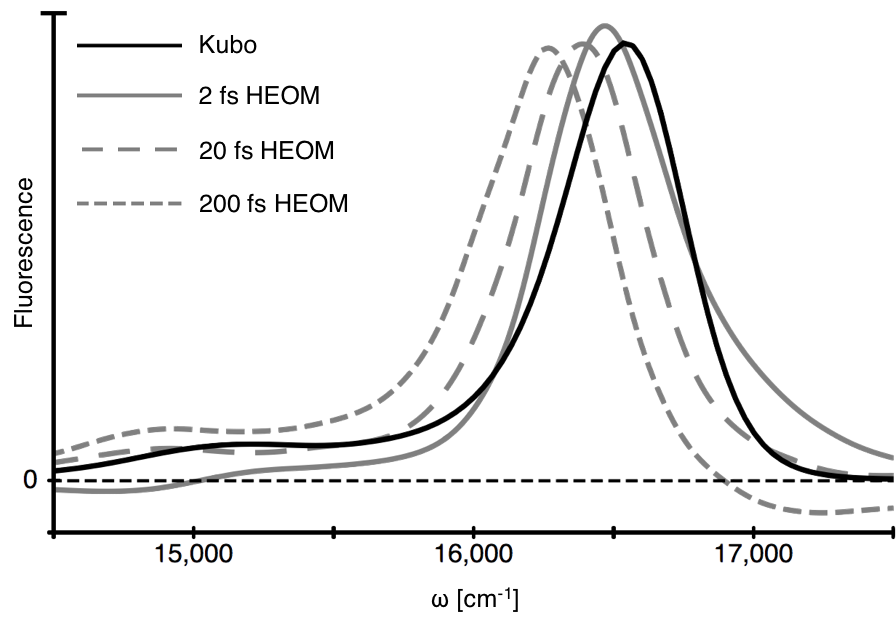}
\caption{\textbf{Time-dependence of HEOM DBV fluorescence compared to Kubo lineshape.} Kubo fluorescence (black) fails to describe the dynamic localization captured by HEOM (gray) over 2 fs (solid), 20 fs (long dashed), and 200 fs (short dashed).}
\end{center}
\end{figure*}

While most of the electronic couplings in PC645 are weak enough (V~$\ll$~50~cm$^{-1}$) that a site-basis description is appropriate, the strong electronic coupling between the DBVs ($\sim$320 cm$^{-1}$) indicates that some exciton delocalization is present. While delocalization can increase coupling, we find that the exciton and site basis description of the DBV core both have at most 50 cm$^{-1}$ of coupling to the PCB82s. In our {\it{ab initio}} QM/MM spectral densities, the DBVs and PCB82s have a long-lived ($\sim$200 fs lifetime), high-frequency ($\sim$1650 cm$^{-1}$) vibration with at most 96 cm$^{-1}$ of reorganization energy which supports a 12 cm$^{-1}$ vibronic coupling. The assignment of $\lambda_{deph}$ for down-conversion in PC645 is complicated by both the presence of strong electronic coupling in the DBV core and the many timescales of vibrational relaxation evidenced by the structured form of the spectral densities. The $\sim$250 cm$^{-1}$ shift in the fluorescence of the DBV core during the first 200 fs following excitation (Figure S10), however, is a lower bound for $\lambda_{deph}$. We thus find that the ratio of $\lambda_{deph}$ to the vibronic coupling is at least 20, indicating the dominance of an incoherent vibronic transport mechanism.

\section*{QM/MM Spectral Density Construction}
\addcontentsline{toc}{section}{QM/MM Spectral Density Construction}
\hspace{\parindent} Starting from the PC645 X-ray structure reported by Curmi et al. \cite{Harrop:2014io}, we constructed AMBER \cite{AMBER2016} molecular mechanics force fields for each bilin with Antechamber, part of AmberTools13 \cite{AMBER2016}. We performed QM/MM nuclear dynamics using NWChem version 6.3 \cite{Valiev:2010bb} with a quantum mechanical treatment of a single bilin in each trajectory. We separated the quantum and classical regions of each calculation by inserting a fictitious hydrogen atom as a link into each bilin-protein bond \cite{Singh:1986cu}. Given the aqueous protein environment, we assumed that carboxylic acid groups were deprotonated, resulting in a net charge of -2 on each bilin. We added all bilin hydrogens by hand while allowing NWChem to automatically add hydrogens to the protein backbone assuming neutral pH. We obtained initial partial charges for bilin force fields from vacuum density functional theory calculations using the B3LYP exchange-correlation functional \cite{Lee:1988fm, Becke:1993is, Stephens:1994jt} and a 6-31G basis set. We solvated the protein using default directives in NWChem. Subsequently, we performed a geometry optimization on the quantum mechanically treated bilin in the presence of protein partial charges prior to running each trajectory. A geometry optimization consisted of ten cycles that each included 3000 MM optimization steps in which we held the QM region fixed followed by ten QM optimization steps in which we held the MM region fixed. Following the geometry optimization, we calculated new partial charges for each bilin in the presence of the protein environment and updated the underlying force field files. After the initial optimization, used to determine the bilin force field parameters, we performed ten additional optimization cycles such that final gradients were below thermal fluctuations. This yielded eight separate optimized structures, one for each bilin, from which to start our QM/MM trajectories.

We began each trajectory with 10 picoseconds of equilibration, during which the simulation temperature stabilized at 295K. We employed Berendsen's thermostat \cite{Berendsen:1984fm}, standard in NWChem, which is known to correctly approximate the canonical ensemble for simulations including thousands of atoms, such as an LHC\cite{Morishita:2000jz}. After equilibration, we performed 40 picosecond production runs. We used a 0.5 femtosecond time step for both equilibration and production runs. Overall, the eight 50 picosecond QM/MM trajectories took nearly nine months to run, and cost over two million CPU hours.

To construct energy gap trajectories for each bilin, we extracted geometries at two femtosecond intervals, for a total of 20,000 geometries per bilin.  We ran time dependent density functional theory (TDDFT) calculations using the B3LYP functional and 6-31G basis set on the resulting 160,000 geometries on both Harvard's Odyssey cluster and the Edison supercomputer at the Department of Energy NERSC facility, using a combined four million CPU hours. We note that while TDDFT often incorrectly predicts excitation energies, cancellation of error results in good descriptions of the curvature of the potential energy surface that we depend on here \cite{Sinnokrot:2001is, Falzon:2005dk}. For each geometry, we calculated the first five excited states and selected the brightest state, defined as the one with the largest oscillator strength. For a small fraction of calculations ($<$1\%) where two states were within 0.05 oscillator strength, we used their transition dipole vectors to find a linear combination of the two states with maximized oscillator strength. Separately, roughly 5\% of the TDDFT calculations failed to converge, requiring us to re-run these calculations with the larger 6-31G* basis set. To make sure that mixing results from two different basis sets did not introduce numerical issues, we additionally ran 6-31G* TDDFT for all geometries of a single bilin. We found the resulting energy gap correlation function to be nearly identical to that constructed with the mixed 6-31G and 6-31G* results.

Following Ref. \cite{Valleau:2012ig}, we constructed two-time bath correlation functions from the energy gap trajectories. To ensure that the correlation functions decayed to zero after roughly two picoseconds, we convolved the resulting correlation functions with a Gaussian having a standard deviation of 24 femtoseconds, as per Ref. \cite{Shim2012a}. We Fourier transformed the resulting correlation functions to obtain a spectral density for each bilin. Drude-Lorentz peaks were then fit to each spectral density to allow for their application to exciton dynamics with the hierarchical equations of motion (HEOM) approach \cite{Tanimura1989a, Tanimura2012a}.

\end{document}